\documentclass[pra,twocolumn,preprintnumbers,amsmath,amssymb,nofootinbib,showpacs]{revtex4-1}
\usepackage{graphicx}
\usepackage{amssymb}
\usepackage{mathrsfs}
\usepackage{color}

\renewcommand\d{\partial}

\newcommand{\mR}{\mathfrak}
\newcommand{\mC}{\mathcal}

\newcommand{\beq}{\begin{equation}}
\newcommand{\eeq}{\end{equation}}

\begin{document}
\preprint{NT@UW-14-18}

\title{Effective theory of two-dimensional chiral superfluids: \\ gauge duality and Newton-Cartan formulation}

\author{Sergej Moroz}
\affiliation{Department of Physics, University of Colorado,
Boulder, Colorado 80309, USA}
\affiliation{Center for Theory of Quantum Matter, University of Colorado, Boulder, Colorado 80309, USA}
\affiliation{Department of Physics, University of Washington,
Seattle, Washington 98195, USA}

\author{Carlos Hoyos}
\affiliation{Raymond and Beverly Sackler Faculty of Exact Sciences,
School of Physics and Astronomy,
Tel-Aviv University, Ramat-Aviv 69978, Israel}


\begin{abstract}
We present a theory of Galilean-invariant conventional and chiral $p_x \pm ip_y$ fermionic superfluids at zero temperature in two spatial dimensions in terms of a dual gauge theory. Our formulation is general coordinate invariant. The parity-violating effects are encoded in the Wen-Zee term that gives rise to the Hall viscosity and edge current. We show that the relativistic superfluid with the Euler current reduces to the chiral superfluid in the limit $c\to\infty$. Using Newton-Cartan geometry we construct the covariant formulation of the effective theory and calculate the energy current.
\end{abstract}


\maketitle

\section{Introduction} \label{intro}

After almost a century since the discovery of  superfluidity in liquid helium, the macroscopic manifestation of quantum mechanics in superfluids is still a fascinating topic of physics \cite{volovik1992exotic,vollhardt,volovikbook}. Modern sophisticated experiments with liquid helium and ultracold atomic gases allow to study various properties of these quantum liquids in great detail. Among different types of superfluids, chiral two-dimensional fermionic superfluids play a prominent role. Originally studied in thin films of $^3\text{He-A}$, nowadays these superfluids attract considerable experimental and theoretical attention in the context of fault-tolerant quantum computation \cite{kitaev2003fault,nayak2008non}.
In this paper we will consider a two-dimensional chiral superfluid with the condensate expressed in momentum space as
\beq \label{porder}
\Delta_{\mathbf{p}}=(p_x\pm i p_y)\hat\Delta,
\eeq
where $\hat\Delta$ is a real function of the magnitude of momentum. Microscopically, this condensate can be realized using spin-polarized (i.e. single-component) fermions with short-range attractive interactions, the system studied before in ultracold experiments \cite{Gunter}. Alternatively, the elusive Moore-Read ($\nu=5/2$) quantum Hall state can be understood as a $p_x\pm ip_y$ superfluid of composite fermions \cite{PhysRevB.61.10267}.

As already realized by Onsager, London and Feynman, the phase of the macroscopic wave function plays a central role in the theory of superfluidity. Today this phase is identified with a gapless Goldstone boson of the broken global particle number symmetry. The low-energy and long-wavelength physics of conventional superfluids can thus be encoded in the effective theory of the Goldstone boson. Interestingly, in two spatial dimensions a $U(1)$ gauge boson (photon) carries just one degree of freedom and has zero spin. This observation suggests the possibility of having a dual description of superfluids  in terms of a gauge field \cite{Fisher}. In Sec. \ref{dualsec} we will realize exactly this idea for Galilean-invariant conventional and chiral superfluids. As will become evident in Sec. \ref{vortex}, it is straightforward to incorporate quantum vortices in the dual description: they are pointlike sources, i.e., charges, of the dual gauge field. Moreover, the duality formulation will allow us in Sec. \ref{rellim} to identify the relativistic theory that gives rise to the chiral superfluid in the nonrelativistic limit $c\to\infty$.

 General coordinate invariance proved to be essential in Einstein's construction of the general theory of relativity. Here we will take advantage of the nonrelativistic version of this principle that was first proposed in \cite{Son:2005rv}. Technically, the effective theory will be invariant under spacetime diffeomorphisms.
 This enables us to study superfluids living on arbitrary two-dimensional spatial manifolds and to use arbitrary space-time coordinates. This formalism is useful even if one is only interested  in flat space physics since it allows to calculate easily various currents and their correlators by taking small variations of the action with respect to external sources. In Sec. \ref{ncG} we will construct the covariant formulation of the theory of superfluids in Newton-Cartan geometry, which appears to be the most natural formalism for nonrelativistic physics \cite{Son2013,Christensen:2013lma,*Christensen:2013rfa,Geracie2014,Gromov2014,Bradlyn2014,Banerjee:2014nja,*Banerjee:2014pya,Brauner2014a}. Using this formulation, we will calculate the energy current for both conventional and chiral superfluids in Sec. \ref{Encur}.

 This paper is a continuation of \cite{Stone2004} and our previous work \cite{Hoyos2013}, where the effective theory of Galilean-invariant chiral superfluids in terms of Goldstone phase was constructed. Our predictions might be relevant for two-dimensional chiral superfluids to be realized in experiments with single-component ultracold fermions.

\section{Dual description of two-dimensional  superfluid} \label{dualsec}
\subsection{Conventional superfluid}
First we consider a conventional nonrelativistic s-wave fermionic superfluid living on some two-dimensional surface with a generically time-dependent metric $g_{ij}$. Since at zero temperature the superfluid does not dissipate energy, it is an isentropic fluid and we can start from the action
\beq \label{act}
S=\int dt d\mathbf{x} \sqrt{g}\mathcal{L}_{sf}
\eeq
with $g=\text{det} g_{ij}$ and the Lagrangian \cite{Zakharov1997, 1998cond.mat..5152S}
\beq \label{HydroLag}
\begin{split}
\mathcal{L}_{sf}=&\frac{1}{2}\rho g_{ij}  v^i v^j-\epsilon(\rho)-\theta \left[ \frac {1}{ \sqrt{g}} \partial_t (\sqrt{g} \rho)+\nabla_i (\rho v^i) \right] \\
                              &-A_t \rho-A_i \rho v^i.
\end{split}
\eeq
Here $v^i$ is the superfluid velocity and $\rho$ is the particle number density,\footnote{In this paper we follow the notation of \cite{Hoyos2013}. This implies that the mass density and the particle number density coincide because we set the mass of the elementary fermion to unity.} $\epsilon$ is the internal energy density and $\nabla_i$ stands for the spatial covariant derivative (Levi-Civita connection). In addition, we included the coupling of the superfluid to the background $U(1)_N$ gauge field $A_\mu$. The term with the Lagrange multiplier $\theta$ ensures the conservation of the particle number. In Appendix \ref{AppA} we demonstrate that $\theta$ is actually the Goldstone field of the broken $U(1)_N$ particle number symmetry. Note that under a constant shift of $\theta$ the action changes only by a total derivative. This is the realization of the $U(1)_N$ symmetry in the effective theory.

In \emph{two} spatial dimensions the $U(1)_N$ particle number current $\mathcal{J}^\mu=(\rho, \rho \mathbf{v})$ can be expressed as\footnote{Here we introduced the tensor $\varepsilon^{\mu\nu\rho}=\frac{1}{\sqrt{g}}\epsilon^{\mu\nu\rho}$, where the totally antisymmetric Levi-Civita symbol is defined by $\epsilon^{tij}\equiv\epsilon^{ij}$ and $\epsilon^{12}\equiv +1$.}
\beq \label{dual}
\mathcal{J}^{\mu}\equiv \varepsilon^{\mu\nu\rho}\partial_\nu a_\rho=\frac{1}{2}\varepsilon^{\mu\nu\rho}f_{\nu \rho},
\eeq
where we introduced the \emph{dual gauge} field $a_\mu$. Indeed, the gauge transformation
\beq \label{dualgauge}
a_\mu \to a_\mu-\partial_\mu \chi
\eeq
leaves the current $\mathcal{J}^\mu$ invariant.
 In the new language, the conservation law of the particle number is
\beq \label{Bianchi}
\epsilon^{\mu\nu\rho}\partial_\mu f_{\nu\rho}=0,
\eeq
which is trivially satisfied.

The transformation \eqref{dual} thus allows us to trade the hydrodynamic theory of the constrained variables $\rho$ and $v^i$  for the theory of the field $a_\mu$ which has the \emph{gauge} freedom. Indeed from the duality relation \eqref{dual} we find
\beq \label{rhov}
\begin{split}
\rho&= b, \\
v^i&=-\frac{\varepsilon^{ij}e_j}{b},
\end{split}
\eeq
where we introduced the dual magnetic field $b\equiv \varepsilon^{ij}\partial_i a_j=g^{-1/2} \epsilon^{ij}\partial_i a_j$ and the dual electric field $e_j\equiv \partial_t a_j-\partial_j a_t$. In the dual language the Lagrangian \eqref{HydroLag} can thus be expressed in the simple form
\beq \label{sfL}
\mathcal{L}_{sf}=\frac{g^{ij} e_i e_j }{2b}-\epsilon(b)- \varepsilon^{\mu\nu\rho}A_{\mu}\partial_{\nu}a_{\rho},
\eeq
which is a \emph{nonlinear} theory of electromagnetism in two spatial dimensions.  As shown in Appendix \ref{AppB}, small Goldstone fluctuations around the homogeneous ground state in flat space with $A_\mu=0$ are described by the linearized version of Eq. \eqref{sfL}, which is just the relativistic Maxwell electrodynamics.

For any effective theory a power counting scheme must be specified that orders various terms according to their importance.
Here we will use the hydrodynamic power counting which allows large Goldstone fluctuations and thus  the velocity and density are not assumed to be small. In other words we set $\rho\sim \mathbf{v}\sim A_\mu \sim \mathcal{O}(1)$ and $\theta\sim \mathcal{O}(p^{-1})$, where $p$ is a small momentum scale. We thus find that $\mathcal{L}_{sf}\sim \mathcal{O}(1)$, i.e., it is of the leading order in the hydrodynamic power counting. For the dual gauge potential this implies  $a\sim \mathcal{O}(p^{-1})$.

The dual theory defined by the Lagrangian \eqref{sfL} is invariant under the nonrelativistic version of  the general coordinate transformations that was introduced in \cite{Son:2005rv}. Indeed, first we observe that with respect to a spatial diffeomorphism $x^i\to x^i +\xi^i (t, \mathbf{x})$, the hydrodynamic fields $\rho$ and $v_i$ transform as follows \cite{Son:2007,Hoyos2013}
\beq
\begin{split}
 \delta\rho&=-\xi^k \partial_k \rho, \\
 \delta v_i &=-\xi^k \partial_k v_i-v_k\partial_i \xi^k+ g_{ik}\dot{\xi}^k.
\end{split}
\eeq
This result together with Eq. \eqref{rhov} implies\footnote{To prove the second equation in \eqref{elmag} we used $\delta v^i=-\xi^k \partial_k v^i+v^k \partial_k \xi^i+\dot\xi^i$, $\delta \sqrt{g}=-\xi^k \partial_k \sqrt{g}-\sqrt{g}\partial_k \xi^k$ and the identity $\varepsilon^{ij}\partial_k \xi^k=\varepsilon^{ik}\partial_k \xi^j+\varepsilon^{kj}\partial_k \xi^i$.}
 \beq \label{elmag}
\begin{split}
\delta b&= -\xi^k \partial_k b, \\
\delta e_i&=-\xi ^k \partial_k e_i-e_k \partial_i \xi^k+b\varepsilon_{ik}\dot \xi^k.
\end{split}
\eeq
We find that these transformation rules are satisfied provided the dual gauge potential transforms simply as a one-form under the spatial general coordinate transformation, i.e.,
\beq
\delta a_{\mu}=-\xi^k \partial_k a_{\mu}-a_k \partial_{\mu}\xi^k.
\eeq
Using this result together with the transformation rules for $A_\mu$ and $g_{ij}$ found in \cite{Son:2005rv}\footnote{The over-dot denotes the temporal derivative.}
\beq \label{trans}
\begin{split}
\delta A_t&=-\xi^k \partial_k A_t-A_k \dot{\xi}^k, \\
\delta A_i&=-\xi^k \partial_k A_i-A_k \partial_i \xi^k+g_{ik}\dot{\xi}^k, \\
\delta g_{ij}&=-\xi^k \partial_k g_{ij}-g_{ik}\partial_j \xi^k-g_{kj}\partial_i \xi^k
\end{split}
\eeq
it straightforward to demonstrate that the Lagrangian \eqref{sfL} transforms as a scalar and the action \eqref{act} is indeed invariant.

Although we do not specify a microscopic fermionic model  and pairing mechanism here, we note that the transformation rules \eqref{trans} are only valid if the ``gyromagnetic ratio'' $\mathrm{g}_\psi$ and the spin $\mathrm{s}_\psi$ of the fermion field in the microscopic model satisfy\footnote{$g_\psi$ is a parameter which introduces a nonminimal coupling of fermions to the background $U(1)_N$ magnetic field of the form $\mathcal{L}_{g}\sim g_\psi B \rho$, where $\rho$ is the superfluid density that coincides with the total density at $T=0$. Since the background gauge field $A_{\mu}$, introduced in this paper, is completely unrelated to the electromagnetic gauge potential, it is important to keep in mind that $g_\psi$ does not coincide with the gyromagnetic ratio of the fermionic atom. In general the value of $g_\psi$ can be determined experimentally by rotating the superfluid (i.e. switching on the $U(1)_N$ magnetic field) and measuring the $U(1)_N$ current.}
\beq \label{gyrs}
\mathrm{g}_\psi-2\mathrm{s}_\psi=0.
\eeq
Throughout this paper we will assume that this relation is valid. Generalization to the case $\mathrm{g}_\psi-2\mathrm{s}_\psi \ne 0$ can be obtained in a straightforward fashion by following \cite{Geracie2014}.

Time reversal and parity transformations are given by
\beq
\begin{split}
T:& \, t\to -t, \, \theta\to -\theta, \, A_i\to -A_i, \, a_t\to -a_t; \\
P:& \,  x_1\leftrightarrow x_2, \, A_1\leftrightarrow A_2, \, a_t\to-a_t, \, a_1\leftrightarrow -a_2.
\end{split}
\eeq
It is now straightforward to check that the Lagrangian \eqref{sfL} is separately invariant under $T$  and $P$.

Notably in the dual formulation we can write the Chern-Simons action which is gauge-invariant and general coordinate invariant \beq
\begin{split}
S_{\text{CS}}=\frac{\nu_a}{4\pi} \int dt d \mathbf{x} \epsilon^{\mu\nu\rho}a_\mu \partial_\nu a_\rho \sim \mathcal{O}(p^{-1}). \\
\end{split}
\eeq
In our power-counting this term is more important than the action  \eqref{act}. In addition, in terms of the original hydrodynamic variables it is nonlocal in position space.
Note, however, that the Chern-Simons term makes the dual photon (aka $U(1)_N$ Goldstone boson) massive \cite{Deser1982} and thus should not appear in the theory of a compressible superfluid. For this reasons in the following we set $\nu_a=0$.

Given a general coordinate invariant theory  it is straightforward  to calculate its stress tensor. General coordinate invariance implies that for $A_k=0$ the contravariant stress tensor can be calculated as \cite{Son:2005rv, Brauner2014}
\beq
T^{ij}= \frac{2}{\sqrt{g}}\frac{\delta S}{\delta g_{ij}}.
\eeq
For the superfluid defined by the Lagrangian \eqref{sfL} we find the ideal fluid result
\beq \label{idstress}
T^{ij}_{\text{\text{ideal}}}=\underbrace{ [\frac{d \epsilon}{db}b-\epsilon] }_{P(b)}g^{ij}+\frac{\mathbf{e}^2 g^{ij}-e^i e^j}{b}=P(\rho) g^{ij}+\rho v^i v^j,
\eeq
where we introduced the pressure $P$ as the function of the superfluid density and used that in two spatial dimensions the projector $\mathbf{e}^2 g^{ij}-e^i e^j=\varepsilon^{ik} e_k \varepsilon^{jl} e_l=\rho^2 v^i v^j$.

In the dual formulation the global $U(1)_N$ particle number symmetry is realized nontrivially.  It is unrelated to the dual gauge symmetry, but appears as the dual magnetic flux symmetry since  the total particle number is given by
\beq
N=\int dt d\mathbf{x} \sqrt{g}b.
\eeq
The flux symmetry is broken spontaneously by the ground state of the dual electrodynamics \cite{Kovner1991}. Under an infinitesimal $U(1)_N$ transformation the sources transform as
\beq
\delta A_\mu =-\partial_\mu \alpha, \quad \delta g_{ij}=0.
\eeq

Finally, it is straightforward to demonstrate that the effective theory is Galilean-invariant for $A_\mu=0$ and $g_{ij}=\delta_{ij}$. The infinitesimal Galilean boost is a combination of the the diffeomorphism $\xi^k=v^k t$ and the gauge transformation $\alpha=v^k x_k$. Galilean transformations are physical symmetries because it does not modify the background fields (see Sec. 2.1 in \cite{Janiszewski:2012nb}).

\subsection{Chiral superfluid}
We now consider a two-dimensional chiral superfluid. In addition to the conventional breaking of the global $U(1)_N$ particle number symmetry it exhibits spontaneous breaking of the spatial rotation symmetry. In the following it will be denoted by $SO(2)_V$, i.e., the group of rotations of the orthonormal two-dimensional vielbein to be introduced below. We will assume the symmetry breaking pattern
\beq
U(1)_N\times SO(2)_V\rightarrow U(1)_D,
\eeq
where $U(1)_D$ stands for the diagonal combination of $U(1)_N$ and $SO(2)_V$ which remains unbroken. As the result at zero temperature the low-energy physics is governed by just \emph{one} Goldstone boson. An important example of such a superfluid is the chiral $p_x\pm i p_y$ fermionic  superfluid briefly introduced in Sec. \ref{intro}. Notably the chiral condensate \eqref{porder} breaks spontaneously time reversal and parity symmetries which gives rise to qualitatively new effects compared to phenomena taking place in the conventional superfluid discussed above.

As a first step towards the dual description of the chiral superfluid we will follow \cite{Hoyos2013} and introduce an orthonormal spatial vielbein $e^a_i$ with $a=1,2$. Since such a vielbein is defined only up to a local $SO(2)_V$ rotation
\beq
e^{a}_{i}\to e^{a}_{i}+\phi(t,\mathbf{x})\epsilon^{ab}e^{b}_{i},
\eeq
we can introduce the spin connection
\beq \label{spincon}
\begin{split}
\omega_t&\equiv\frac{1}{2}\Big(\epsilon^{ab}e^{aj} \partial_t e^{b}_{j}+B \Big), \\
\omega_i&\equiv\frac{1}{2}\epsilon^{ab}e^{aj} \nabla_i e^{b}_{j}=\frac{1}{2}\Big(\epsilon^{ab}e^{aj} \partial_i e^{b}_{j}-\varepsilon^{jk}\partial_j g_{ik} \Big),
\end{split}
\eeq
where we defined  $e^{aj}\equiv e^{a}_{i}g^{ij}$ and the magnetic field $B\equiv\varepsilon^{ij}\partial_i A_j$. By construction, under a local $SO(2)_V$ rotation the connection transforms as an abelian gauge field, i.e.,
\beq
\omega_{\nu}\to \omega_{\nu}-\partial_{\nu} \phi.
\eeq
In addition, under spatial diffeomorphisms $\omega_\nu$ transforms simply as a one-form
\beq
\delta \omega_{\mu}=-\xi^k \partial_k \omega_{\mu}-\omega_k \partial_{\mu}\xi^k.
\eeq
Note that under the discrete symmetries $\omega_\nu$ transforms similar to the dual gauge field $a_\mu$
\beq
\begin{split}
T:& \,  \omega_t\to -\omega_t; \\
P:& \,  \omega_t\to-\omega_t, \, \omega_1\leftrightarrow -\omega_2.
\end{split}
\eeq

The dual effective theory of the chiral superfluid is now obtained by adding to the Lagrangian \eqref{sfL}
the general coordinate invariant  Wen-Zee term \cite{Wen1992}
\beq
\begin{split}
\mathcal{L}_{WZ}&=-s \varepsilon^{\mu\nu\rho}\omega_{\mu}\partial_\nu a_\rho \\
&=-s\rho(\omega_t+\omega_i v^i).
\end{split}
\eeq
Within our power counting this sub-leading term is of order $O(p)$. Provided the parameter $s$ is kept fixed, $\mathcal{L}_{WZ}$ breaks separately parity and time reversal, but preserves  the combined $PT$ symmetry. If one transforms the chirality of the ground state $s\to -s$, both $P$ and $T$ are preserved separately by the Wen-Zee term. One must set $s=\pm 1/2$ for the $p_x\pm i p_y$ superfluid.\footnote{The effective theory can also be used for the low-energy description of anyon superfluids. In particular, for anyons with the statistical phase angle $\theta=\pi(1-1/n)$ \cite{Greiter:1989qb} we must fix $s=(n-1/n)/2$. Similar to the chiral superfluid, our construction is valid only if the ``gyromagnetic ratio'' and the spin of the anyon are fine-tuned to satisfy Eq. \eqref{gyrs}. This restriction can be easily relaxed by following arguments of \cite{Geracie2014}.} As has been realized recently in \cite{Tada2014,Volovik2014,Huang2014}, the two-dimensional chiral pairing in higher partial waves is more subtle. We defer the construction of the effective theory for this case to a future work. 

It is well-known that the chiral superfluid studied here is a topological quantum liquid since its ground state has different topological properties in the weakly (BCS) and strongly (BEC) coupled regimes that are separated by a quantum phase transition \cite{volovikbook, PhysRevB.61.10267}. This implies the presence of a protected gapless fermionic Majorana mode localized on the boundary between the two phases. We emphasize that in our construction we did not specify the equation of state $\epsilon(\rho)$ and thus the effective theory described here should be valid in both phases. Although the Majorana mode does not appear as an explicit degree of freedom, it is integrated out and gives rise to nonanalyticity of the term $\epsilon(b)$ in the Lagrangian at the phase transition point.

Since in the effective theory of chiral superfluids the spatial vielbein does not appear linearly, but only quadratically, it is natural to expect that  the introduction of the vielbein and spin connection is actually not necessary and that the theory can be formulated covariantly using the spatial metric $g_{ij}$ only. While this is not obvious within the formalism presented in \cite{Hoyos2013}, it is straightforward to eliminate the vielbein in the dual formalism developed here. Indeed, up to a surface term we can rewrite the Wen-Zee Lagrangian as
\beq \label{WZ}
\mathcal{L}_{WZ}=-s \varepsilon^{\mu\nu\rho}a_{\mu}\partial_\nu \omega_\rho=-s\left(a_t B_\omega-\varepsilon^{ij}a_i E_{\omega j} \right),
\eeq
where we introduced the gravitomagnetic field $B_\omega\equiv\varepsilon^{ij}\partial_i \omega_j$ and the gravitoelectric field $E_{\omega j}\equiv\partial_t \omega_j-\partial_j \omega_t$. In Appendix \ref{AppC} we show that
\beq
\begin{split}
B_\omega&=\frac{1}{2}R, \\
E_{\omega i}&=\frac{1}{2}\left[-\partial_t (\Gamma^k_{ij})\varepsilon^{jl} g_{kl}-\partial_i B \right],
\end{split}
\eeq
where $R$ and $\Gamma^{k}_{ij}$ stand for the Ricci scalar and Chirstoffel symbol respectively.  Thus the Wen-Zee term can be indeed written only in terms of the metric $g_{ij}$ and its derivatives.

The Wen-Zee term gives rise to novel phenomena. In the context of quantum Hall effect these were investigated for example in \cite{Wen1992, Hoyos:2011ez,2014arXiv1403.5809G}. Here we study its consequences for the chiral superfluid. First, due to the presence of the magnetic field $B$ in Eq. \eqref{spincon}, it leads to the modification of the $U(1)_N$ current
\beq
\mathcal{J}^{\mu}_{\text{edge}}=-\frac{1}{\sqrt{g}}\frac{\delta S_{WZ}}{\delta A_{\mu}}=(0, \frac{s}{2}\varepsilon^{ij}\partial_j \rho).
\eeq
This is the well-known Mermin-Muzikar edge current  \cite{Mermin1980} responsible for the macroscopic angular momentum of the chiral ground state
\beq \label{angmom}
L_{\text{GS}}=\int d^2x \epsilon_{ij}x^i \mathcal{J}^j_{\text{edge}}=s\int d^2x \rho.
\eeq
Second, the stress tensor is also modified compared with the ideal fluid result \eqref{idstress}.
Indeed from the variation of the connection under a small variation of the metric \cite{Hoyos2013}
\beq
\begin{split}
\delta \omega_t=&-\frac{1}{4}\varepsilon^{in}g^{jk}\partial_t g_{nk}\delta g_{ij} -\frac{1}{4}B g^{ij}\delta g_{ij}, \\
 \delta \omega_l=&-\frac{1}{4}\varepsilon^{in}g^{jk}\partial_l g_{nk}\delta g_{ij}-\frac{1}{2}\varepsilon^{jk}\partial_j \delta g_{lk}\\
 &+\frac{1}{4}\varepsilon^{mk}\partial_m g_{lk} g^{ij}\delta g_{ij}
\end{split}
\eeq
we find
\beq
\delta S_{WZ}=-s \int dt d\mathbf{x} \epsilon^{\mu\nu\rho}\delta \omega_\mu \partial_\nu a_\rho
\eeq
which gives rise to the modification\footnote{We  used the Gauss law \eqref{Gauss} to obtain the term $\sim s^2$ in Eq. \eqref{stress}.}
\beq \label{stress}
\begin{split}
\Delta T^{ij}_{WZ}&= \frac{2}{\sqrt{g}}\frac{\delta S_{WZ}}{\delta g_{ij}}\\
         &=(v^i \mathcal{J}^j_{\text{edge}}+v^j \mathcal{J}^i_{\text{edge}})+T^{ij}_{\text{Hall}}-\frac{s^2}{4}\rho R g^{ij},
\end{split}
\eeq
where we introduced the Hall viscosity stress tensor \cite{Avron:1995fg,1997physics..12050A,Hoyos2014}
\beq
T^{ij}_{\text{Hall}}=-\eta_H (\varepsilon^{ik}g^{jl}+\varepsilon^{jk}g^{il}) V_{kl}
\eeq
with $V_{kl}\equiv \frac{1}{2}\left(\nabla_k v_l+\nabla_l v_k+\partial_t g_{kl} \right)$ and $\eta_H=-\frac{s}{2}\rho$.
In summary, the dual Wen-Zee term leads to the parity and time reversal violating effects such as the edge current and the Hall viscosity. For a detailed discussion of these effects we refer to \cite{Hoyos2013,Hoyos2014}.

Now that we have the stress tensor, it is straightforward to demonstrate that the invariance of the action under a small spatial diffeomorphism $\xi^i$
\beq
S[a_\mu+\delta a_\mu, A_{\nu}+\delta A_{\nu}, g_{ij}+\delta g_{ij}
]=S[a_\mu, A_{\nu}, g_{ij}
],
\eeq
implies the Euler equation
\beq \label{Eulereq}
\frac{1}{\sqrt{g}}\partial_t (\sqrt{g}J_k)+\nabla_i T^{i}_{\, k}=E_k J^t+\varepsilon_{ik}J^i B,
\eeq
where we introduced $T^{i}_{\, k}\equiv T^{ij}g_{jk}$ and the total $U(1)_N$ current $J^\mu\equiv\mathcal{J}^\mu+\mathcal{J}^\mu_{\text{edge}}$.

Finally, it is instructive to demonstrate how the hydrodynamic conservation equations arise in the dual formalism. As noted above, the conservation of the particle density is simply encoded in the Bianchi identity \eqref{Bianchi}. On the other hand, the Euler-Lagrange equations give rise to the vorticity and hydrodynamic Euler equations. Indeed, since the daul gauge field always appears with a derivative, the equations of motion are given by
\beq \label{FE}
\partial_\mu \left[ \sqrt{g} \frac{\partial \mathcal{L}_{ch}}{\partial \partial_\mu a_\nu} \right]=0,
\eeq
where $\mathcal{L}_{ch}=\mathcal{L}_{sf}+\mathcal{L}_{WZ}$.
The Gauss law ($\nu=t$) reads
\beq
\frac{1}{\sqrt{g}}\partial_i\left(\sqrt{g}\frac{e^i}{b}\right)=B+\frac{s}{2}R
\eeq
or in the covariant form
\beq \label{Gauss}
g^{ij}\nabla_i\frac{e_j}{b}=B+\frac{s}{2}R.
\eeq
The external magnetic field and Ricci curvature play the role of a background smooth charge distribution for the dual gauge field. If we define the vorticity
$
W \equiv \frac 1 2 \varepsilon^{ij}\nabla_i v_j=\frac 1 2 \nabla_i  \frac {e^i}{b},
$
the Gauss law becomes the vorticity equation
\beq \label{vorticity}
2W=B+ \frac s 2 R.
\eeq
It is straightforward to check that the spatial components ($\nu=k$) of Eq. \eqref{FE} give rise to the Euler equation. In terms of the hydrodynamic variables it is given by
\beq
D_t v_k+\frac{\nabla_k P}{\rho} =E_k+s E_{\omega k}+\left(B+s B_\omega\right)\varepsilon_{ik}v^i,
\eeq
where we introduced the material derivative $D_t\equiv \partial_t+v \cdot \nabla$ and used the Gauss equation \eqref{Gauss}. Although it is not manifest, this equation is equivalent to Eq. \eqref{Eulereq}.

\section{Vortices} \label{vortex}
It is evident from Eq. \eqref{vorticity} that, in the presence of a background magnetic field $B$, a two-dimensional superfluid carries vorticity. Moreover, in a chiral superfluid the vorticity is also sourced by the Ricci curvature $R$ of two-dimensional space. While any regular superfluid flow is necessarily irrotational, the vorticity in a superfluid originates from singular solutions known as quantum vortices. Due to conservation of the topological winding number in a \emph{static} background, the total number of vortices is strictly conserved in that case. Locally this leads to the conservation law
\beq
\frac {1}{ \sqrt{g}} \partial_t (\sqrt{g} J^t_v)+\nabla_i J^i_v=0,
\eeq
where we introduced the topological vortex current $J^\mu_v$. In terms of the Goldstone field $\theta$ this current is given by\footnote{The prefactor $1/\pi$ in Eq. \eqref{vortexc} appears in the case of a fermionic superfluid. Note that for a bosonic superfluid the prefactor is two times smaller.}
\beq \label{vortexc}
J^\mu_{v}=\frac 1 \pi \varepsilon^{\mu\nu\rho} \partial_\nu \partial_\rho \theta,
\eeq
which implies $J_v\sim \mathcal{O}(p)$ in our power counting scheme.

Consider a superfluid living on a closed spatial manifold $\mathcal{M}$. Since in a fermionic superfluid a vortex carries $\pi/2$ units of vorticity, we can use Eq. \eqref{vorticity} and find that the total number of vortices is given by
\beq \label{shift}
N_v=\int d \mathbf{x} \sqrt{g} J^0_v=\frac \Phi \pi+2s \chi,
\eeq
where $\Phi=\int d \mathbf{x} \sqrt{g} B$ is the total magnetic flux piercing $\mathcal{M}$ and the Euler characteristic $\chi=2-2g$, where $g$ is the genus of $\mathcal{M}$.  This is the reason why for the chiral superfluid the total number of vortices is sensitive to the topology of the manifold $\mathcal{M}$. For example, for a $p_x\pm i p_y$ superfluid on a sphere in the absence of magnetic flux one finds $N_v=\pm 2$. The formula \eqref{shift} is analogous to the one in the quantum Hall effect, with the second term known as the shift \cite{Wen1992}.

Now we will extend our effective theory by adding the vortex part to the dual Lagrangian. This is relevant if vortices are present in the ground state, which happens if the right-hand-side of Eq. \eqref{shift} is nonvanishing. Since the vortex current is conserved it can be dualized
\beq
J^\mu_{v}= \varepsilon^{\mu\nu\rho} \partial_\nu b_\rho,
\eeq
where we introduced the gauge field $b_{\rho}\sim \mathcal{O}(1)$ dual to the vortex current. This field transforms as a one-form under general coordinate transformations.
Up to the next-to-leading order in our power counting we can now generically add to the Lagrangian the following terms\footnote{Note that an additional general coordinate invariant term $f(b)\varepsilon^{ij}\partial_i b_j$ can be eliminated from the vortex Lagrangian by the redefinition $a_\mu\to a_\mu+\sigma(b) b_\mu$ with a properly chosen function $\sigma(b)$.} that are general coordinate invariant
\beq
\begin{split}
\mathcal{L}_v=&-\tilde q_v \varepsilon^{\mu\nu\rho}a_\mu \partial_\nu b_\rho- q_v \left(\varepsilon^{\mu\nu\rho}A_\mu \partial_\nu b_\rho -\frac{g^{ij}e^{\text{v}}_i e^{\text{v}}_j}{2b^{\text{v}}}\right) \\
&+\frac {\nu_b}{4\pi} \varepsilon^{\mu\nu\rho}b_\mu \partial_\nu b_\rho,
\end{split}
\eeq
where  we defined $b^{\text{v}}\equiv \varepsilon^{ij}\partial_i b_j$ and $e^{\text{v}}_j\equiv \partial_t b_j-\partial_j b_t$. Our normalization of the current \eqref{vortexc} also implies $\tilde q_v=-\pi$.  The first two terms make the vortex charged with respect to $a_{\mu}$ and $A_{\mu}$ respectively, while the third term transmutes its quantum statistics \cite{Wilczek1983}. It is well-known that  in a conventional two-dimensional superfluid the vortex is a point-like \emph{boson} that is charged with respect to the dual gauge field $a_\mu$, but is neutral with respect to $A_{\mu}$ \cite{Haldane1985, Lee1991}. For this reason $q_v=\nu_b=0$ for the conventional superfluid. On the other hand, vortices are known to be abelian anyons in a chiral superfluid \cite{Ariad2014}, which implies $\nu_b\ne 0$. In addition, in the weakly coupled BCS phase they accommodate gappless Majorana fermionic modes which can be included in the effective theory  \cite{Hansson2012,Hansson2013}.
We defer better understanding of of the vortex physics in the BCS phase in chiral superfluids to a future work.
\section{Relativistic superfluid and nonrelativistic limit} \label{rellim}
In this section we demonstrate that one can obtain the chiral superfluid as the nonrelativistic limit ($c\to\infty$) of the relativistic superfluid found recently in \cite{GRS}. A similar procedure was used in  \cite{Andreev:2013qsa,*Andreev:2014gia} to derive nonrelativistic invariant actions for Hall systems.

Here we briefly review the construction of \cite{GRS}. The relativistic theory is formulated in the dual language, where the relativistic $U(1)$ current
\beq
\mR{j}^\mu=\mR{n}\mR{u}^{\mu}=\varepsilon^{\mu\nu\rho}\partial_\nu \mR{a}_\rho.
\eeq
Here the relativistic dual gauge field $\mR{a}_{\mu}\sim \mathcal{O}(p^{-1})$ was introduced. The three-velocity satisfies $\mR{u}^{\mu}\mR{u}_\mu=-1$. The effective theory is defined by the gauge invariant action
\beq \label{actrel}
S=\int d^3x \sqrt{|\mR{g}|}\big( \underbrace{\mathcal{L}_0}_{ \mathcal{O}(1)}+ \underbrace{\mathcal{L}_1}_{ \mathcal{O}(p)}\big),
\eeq
where $\mR{g}_{\mu\nu}$ is the spacetime metric and $\mR{g}\equiv\det \mR{g}_{\mu\nu}$. The leading order Lagrangian is  given by
\beq
\mathcal{L}_0=-\epsilon^{rel}(\mR{n})/c-\varepsilon^{\mu\nu\lambda}\mR{A}_\mu \partial_\nu \mR{a}_\lambda,
\eeq
where $\epsilon^{rel}(\mR{n})$ is the relativistic energy density as the function of
 $\mR{n}= \sqrt{\mR{f}_{\mu\nu}\mR{f}^{\mu\nu}/2}$ ($\mR{f}_{\mu\nu}=\partial_\mu \mR{a}_\nu-\partial_\nu \mR{a}_\mu$), and $\mR{A}_\mu$ is the background $U(1)$ gauge field.

Up to redefinitions the subleading part of the Lagrangian can be written as\footnote{One may argue that additional terms are allowed. For example, $f(n)\varepsilon^{\mu\nu\lambda}\mR{u}_\mu \partial_\nu \mR{u}_\lambda\sim \mathcal{O}(p)$ should be included into $\mathcal{L}_1$.  This term, however, can be eliminated by the redefinition $\mR{a}_\mu\to \mR{a}_\mu+\chi(\mR{n})\mR{u}_\mu$ with the properly chosen function $\chi(\mR{n})$. Our choice of $\chi(\mR{n})$ differs from \cite{GRS}, where it was chosen to  eliminate the term $\xi(\mR{n})\mR{F}_{\mu\nu}\mR{f}^{\mu\nu}$ but keep $f(n)\varepsilon^{\mu\nu\lambda}\mR{u}_\mu \partial_\nu \mR{u}_\lambda$.}
\beq \label{L1}
\mathcal{L}_1=\xi(\mR{n})\mR{F}_{\mu\nu}\mR{f}^{\mu\nu}+\kappa\mR{a}_\mu \mR{J}^\mu,
\eeq
where $\mR{F}_{\mu\nu}$ is the field strength of the $U(1)$ gauge field $\mR{A}_\mu$
and the Euler topological current is
\beq \label{Euler}
\mR{J}^\mu= \frac{1}{8\pi} \varepsilon^{\mu\nu\lambda} \varepsilon^{\alpha\beta\gamma} \mR{u}_\alpha \left(\nabla_\nu \mR{u}_\beta \nabla_\lambda \mR{u}_\gamma-\frac{1}{2} R_{\nu \lambda \beta \gamma}   \right).
\eeq
As explained in \cite{GRS}, the Euler current is identically conserved, i.e.,  $\nabla_\mu \mR{J}^\mu=0$. While $\xi(\mR{n})$ can not be fixed by a symmetry argument only, the gauge invariance requires $\kappa$ to be a constant. For a detailed discussion of the Euler current and the effective theory of the relativistic superfluid we refer the reader to \cite{GRS}.

First, we perform the nonrelativistic limit for the conventional superfluid defined by the leading-order Lagrangian $\mathcal{L}_0$. To this end we use
\beq
\begin{split}
&x^\mu=(ct, \, x^i), \qquad \partial_\mu=(c^{-1}\partial_t, \, \partial_i), \\
&\mR{A}_\mu=(c^{-1}\mR{A}_t, \, \mR{A}_i), \qquad \mR{a}_\mu=(c^{-1}\mR{a}_t, \, \mR{a}_i)
\end{split}
\eeq
and decompose the relativistic energy density into the rest mass part and the internal part, i.e.,
\beq
\epsilon^{rel}=\mR{n}c^2+\epsilon.
\eeq
In addition, in the nonrelativistic regime it is convenient to parametrize the metric in terms of $g_{ij}$, $\mC{A}_i$ and $\mC{A}_t$ \cite{Son:2005rv}
\beq \label{metric}
\mR{g}_{\mu\nu}=\left(
\begin{tabular}{ cc }
 $ -1-\frac{2 \mC{A}_t} {c^2}$ & $-\frac{\mC{A}_i}{c}$  \\
  $-\frac{\mC{A}_i}{c}$ & $g_{ij}$
\end{tabular}
 \right).
\eeq
Relativistic covariance implies that $\mC{A}_\mu$ transforms as the gauge field in Eq. \eqref{trans} under spatial diffeomorphisms.

Now using
\beq
\begin{split}
|\mR{g}|&=\left[1+\frac{2 \mC{A}_t+\mC{A}_i \mC{A}^i}{c^2}\right] g+\mC{O}(1/c^4), \\
\mR{n}&=b-\frac{e^2}{2bc^2}-\frac{\varepsilon^{ij}\mC{A}_i e_j}{c^2}-\frac{\mC{A}_i \mC{A}^i b}{2c^2}+\mC{O}(1/c^4)
\end{split}
\eeq
we arrive at
\beq \label{S0}
S_0=-c^2\int dt d\mathbf{x} \sqrt{g} b+ \int dt d\mathbf{x} \sqrt{g}\mathcal{L}_{sf}+\mathcal{O}(1/c^2).
\eeq
Here $\mathcal{L}_{sf}$ is given by Eq. \eqref{sfL}, where we identified the $U(1)_N$ gauge potential $A_\mu=\mR{A}_\mu+\mC{A}_\mu$. After subtracting the rest mass term\footnote{The first term in \eqref{S0} can be removed by adding an appropriate chemical potential. This is achieved by shifting $\mR{A}_0$ as  $\mR{A}_0=-c+ \mR{A}_t/c$.}  from $S_0$ one recovers the nonrelativistic theory \eqref{sfL} describing the conventional superfluid.
 Notably the action depends only on the linear combination $\mR{A}_\mu+\mC{A}_\mu$, but not on $\mR{A}_\mu$ and $\mC{A}_\mu$ separately. Physically this means that the momentum density must be proportional to the particle number current, which within our conventions can be written simply as
\beq \label{cond}
J^i=T^{0i}.
\eeq
This is not a surprise since this result is valid for $A_i=0$ in any general coordinate invariant system composed of single species of particles provided Eq. \eqref{gyrs} is fulfilled \cite{Son:2005rv, Brauner2014, Geracie2014}.

Now we perform the nonrelativistic limit of the sub-leading Lagrangian $\mathcal{L}_1$. As demonstrated in Appendix \ref{AppD}, in this limit the Euler current is given by
\beq
\begin{split}
\mR{J}^0&=\frac{B_\omega}{4\pi}+\mathcal{O}(1/c^2), \\
\mR{J}^i&=-\frac{\varepsilon^{ij}(E_{\omega j}+\partial_j \mR{B}/2)}{4\pi c}+\mathcal{O}(1/c^3),
\end{split}
\eeq
where we introduced the magnetic field constructed from $\mR{A}_i$, i.e., $\mR{B}=\varepsilon^{ij}\partial_i \mR{A}_j$. As a result, we find
\beq
\begin{split}
S_1=&\frac{\kappa}{4\pi} \int dt d\mathbf{x} \sqrt{g} \left(a_t B_\omega-\varepsilon^{ij}a_i \{ E_{\omega j}+\partial_j \mR{B}/2 \}  \right)+ \\
&2c \int dt d\mathbf{x} \sqrt{g}  \xi(b) b \mR{B}+\mathcal{O}(1/c).
\end{split}
\eeq
In the following we will assume $\xi(b)\sim 1/c$, which leads to a finite nonrelativistic limit. Moreover, the requirement \eqref{cond} fixes $\xi(b)$ to be
\beq \label{cond1}
\xi(b)=\frac{\kappa}{8\pi c}+\mathcal{O}(1/c^3)
\eeq
leading finally to
\beq
S_1=\int dt d\mathbf{x} \sqrt{g} \mathcal{L}_{WZ}+\mathcal{O}(1/c^2)
\eeq
with $s=-\kappa/4\pi$. This proves that the relativistic superfluid defined by Eqs. \eqref{actrel}-\eqref{L1} reduces to the chiral superfluid is the nonrelativistic limit.

Finally we must emphasize that the condition \eqref{cond1} is a direct consequence of Eq. \eqref{gyrs} which is assumed to be true throughout this paper. For $\mathrm{g}_\psi-2\mathrm{s}_\psi \ne 0$ one must fix $\xi(b)$ differently since in that case Eq. \eqref{cond1} is generalized to \cite{Geracie2014}
\beq
J^i=T^{0i}-\frac{\mathrm{g}_\psi-2\mathrm{s}_\psi}{2} \varepsilon^{ij}\partial_j J^0.
\eeq
For example, if one sets $\mathrm{g}_\psi-2\mathrm{s}_\psi=-\kappa/(4\pi)$ then $\xi(b)$ must vanish in the nonrelativistic limit, i.e., $\xi(b)=\mathcal{O}(1/c^3)$.
\section{Newton-Cartan formalism} \label{ncG}
So far we imposed nonrelativistic general coordinate invariance only under \emph{spatial} diffeomorphisms $\xi^i(t, \mathbf{x})$. It is possible to include also the symmetry under \emph{temporal} diffeomorphisms $\xi^t(t,\mathbf{x})$ which generate a local reparametrization of time. This extended version of nonrelativistic  general coordinate invariance under $\xi^\mu(t,\mathbf{x})$ with $\mu=(t,i)$ was first demonstrated to be valid for a theory of nonrelativistic particles with no interactions \cite{Son2008} and more recently for the theory of fractional quantum Hall effect \cite{Geracie2014} (see also  \cite{Christensen:2013lma,*Christensen:2013rfa,Gromov2014,Bradlyn2014,Banerjee:2014nja,*Banerjee:2014pya,Brauner2014a}).
Here we will \emph{assume} the invariance of the effective theory of superfluids with respect to temporal and spatial diffeomorphisms, which leads to the following transformation rule for the background fields \cite{Son2008,Geracie2014}
\beq \label{new-gci}
\begin{split}
  \delta A_t& = -\xi^\mu\partial_\mu A_t - A_\mu \dot\xi^\mu  , \\
  \delta A_i &=-\xi^\mu\partial_\mu A_i- A_\mu \partial_i \xi^\mu  + e^\Phi g_{ij}\dot\xi^j
      ,\\
  \delta\Phi &= -\xi^\mu\partial_\mu\Phi+ \dot \xi^t - C_i\dot \xi^i,\\
  \delta C_i &=-\xi^\mu\partial_\mu C_i - C_j \partial_i \xi^j+ \partial_i\xi^t + C_i(\dot\xi^t - C_j\dot\xi^j)
      ,\\
  \delta g_{ij} &=-\xi^\mu\partial_\mu g_{ij}
    -g_{kj}\partial_i\xi^k - g_{ik}\partial_j\xi^k -(C_ig_{jk}+C_jg_{ik})\dot\xi^k, 
 \end{split}
\eeq
where $C^i\equiv g^{ij}C_j$ and $\xi^\mu \partial_\mu\equiv \xi^t \partial_t+\xi^i \partial_i$. This is a generalization of Eq. \eqref{trans}. Here we introduced two additional background fields $\Phi$ and $C_i$ which couple to the energy density and current respectively (see Sec. \ref{Encur} for more details).

The transformation rules \eqref{new-gci} for $\Phi$, $C_i$ and $g_{ij}$ follow most naturally from Newton-Cartan geometry which was developed by Cartan with intention to geometrize Newtonian gravity. We will briefly review its basics here and refer the reader to  \cite{Son2013,Christensen:2013lma,*Christensen:2013rfa,Geracie2014,Gromov2014,Bradlyn2014,Banerjee:2014nja,*Banerjee:2014pya,Brauner2014a} for a detailed presentation.  Subsequently, the covariant formulation of the effective theory of the conventional and chiral superfluid will be presented in Newton-Cartan spacetime.

\subsection{Geometry}

A Newton-Cartan spacetime is a manifold that comes with a degenerate metric tensor with upper indices $g_{\text{nc}}^{\mu\nu}$, a one-form $n_\mu$ and a velocity vector $V^\mu$ with the properties
\beq
n_\mu g_{\text{nc}}^{\mu\nu}=0, \qquad n_\mu V^\mu=1.
\eeq
Given ($g_{\text{nc}}^{\mu\nu}$, $n_\mu$,  $V^\mu$), we can uniquely introduce the metric tensor with lower indices $g^{\text{nc}}_{\mu\nu}$ by imposing the conditions
\beq
g_{\text{nc}}^{\mu\rho} g^{\text{nc}}_{\rho\nu}=\delta^\mu_\nu-V^\mu n_\nu, \qquad g^{\text{nc}}_{\mu\nu}V^\nu=0.
\eeq
Now we can define a connection
\beq
\Gamma^{\rho}_{\, \mu\nu}\equiv V^\rho \partial_\mu n_\nu +\frac{1}{2}g_{\text{nc}}^{\rho \sigma} \left( \partial_\mu g^{\text{nc}}_{\nu\sigma}+\partial_\nu g^{\text{nc}}_{\mu \sigma}-\partial_\sigma g^{\text{nc}}_{\mu\nu} \right)
\eeq
in Newton-Cartan spacetime.
Notably, the connection is not symmetric in the lower indices which gives rise to the nontrivial torsion tensor
\beq
T^\rho_{\mu\nu}\equiv 2\Gamma^\rho_{\, [\mu\nu]}=2V^\rho \partial_{[\mu}n_{\nu]}.
\eeq
Obviously, the torsion vanishes provided the form $n_\mu$ is closed, i.e., $dn=\partial_{[\mu} n_{\nu]}=0$. Here we will impose a weaker condition, namely $n\wedge d n=n_{[\mu}\partial_\nu n_{\rho]}=0$ which insures an absolute notion of space. This follows from Frobenius theorem because in this case there is a unique spatial slicing of Newton-Cartan spacetime which $n_\mu$ is normal to. We mention that in the language of \cite{Christensen:2013lma,*Christensen:2013rfa,Gromov2014,Bradlyn2014,Brauner2014a} the torsion considered in this paper is purely temporal. A more general Newton-Cartan geometry with spatial torsion was discussed in \cite{Christensen:2013lma,*Christensen:2013rfa,Bradlyn2014,Brauner2014a}.

To make connection with the transformation law \eqref{new-gci} we use the following parametrization \cite{Geracie2014}
\beq \label{nV}
n_\mu=\left(
\begin{tabular}{ cc}
 $ e^{-\Phi}$, &
  $-e^{-\Phi} C_i$
\end{tabular}
 \right),
\qquad
V^\mu=\left(
\begin{tabular}{ c}
 $ e^\Phi(1+C_j v^j)$ \\
  $e^\Phi v^i$
\end{tabular}
 \right),
\eeq
which is consistent with $n \cdot V=1$. Since $n_\mu$ and $V^\mu$ transform in Newton-Cartan spacetime simply as
\beq \label{nVtrans}
\begin{split}
\delta n_\mu& = -\xi^\kappa \partial_\kappa n_\mu-n_\kappa \partial_\mu \xi^\kappa, \\
\delta V^\mu& = -\xi^\kappa \partial_\kappa V^\mu+V^\kappa \partial_\kappa \xi^\mu,
\end{split}
\eeq
we can easily reproduce the last three equations in \eqref{new-gci} with the help of Eq. \eqref{nV}. In addition, the parametrization \eqref{nV} implies the following expressions for the metric tensor
\beq \label{metricnc}
\begin{split}
g^{\text{nc}}_{\mu\nu}&=\left(
\begin{tabular}{ cc}
 $v^2$ & $-v_j-v^2 C_j$ \\
 $-v_i-v^2 C_i$ & $g_{ij}+v_i C_j+v_j C_i+v^2 C_i C_j$ \\
\end{tabular}
 \right), \\
g_{\text{nc}}^{\mu\nu}&=\left(
\begin{tabular}{ cc}
 $C^2$ & $C^j$ \\
 $C^i$ & $g^{ij}$ \\
\end{tabular}
 \right).
 \end{split}
\eeq
Finally, we notice that the condition $n\wedge d n=n_{[\mu}\partial_\nu n_{\rho]}=0$ leads to the constraint on the source $C_i$
\beq \label{const}
\epsilon^{ij}[\partial_i C_j+C_i \partial_t C_j]=0.
\eeq

While $A_\mu$ does not transform as a one-form under nonrelativistic general coordinate transformations, we can modify it as follows \cite{Geracie2014}
\beq
\begin{split} \label{Atilde}
  \tilde A_t &\equiv A_t + \frac{1}{2} e^\Phi g_{ij} v^i v^j, \\
  \tilde A_i &\equiv A_i - e^\Phi g_{ij} v^j-\frac 1 2 e^\Phi g_{kl} v^k v^l C_i.
\end{split}
\eeq
A simple derivation of Eq. \eqref{Atilde} can be found in \cite{Brauner2014a}.
Using Eqs. \eqref{nV} and \eqref{nVtrans}, one can check that $\tilde A_\mu$ transforms as a one-form, i.e.,
\beq
\delta \tilde A_{\mu}=-\xi^\kappa \partial_\kappa \tilde A_{\mu}-\tilde A_\kappa \partial_\mu \xi^\kappa.
\eeq

In the following we will need a spin connection in Newton-Cartan geometry. Within Newton-Cartan formalism it is given by
\beq \label{omega}
\omega_\mu=\frac 1 2 \epsilon^{ab} e^{a\nu}\nabla^{nc}_\mu e^b_\nu,
\eeq
where $\nabla^{nc}_\mu$ stands for the covariant derivative in Newton-Cartan spacetime and $e^a_\mu$ denotes the vielbein with $a=1,2$.
For $\Phi=C_i=0$ the components of $\omega_\mu$ were calculated in \cite{Hoyos2013, Son2013}.
It is straightforward to generalize the construction to the case of non-vanishing $\Phi$ and $C_i$. Indeed, in this case $e^a_\mu$ can be parametrized using the spatial vielbein $e^a_i$, spatial velocity $v^i$ and the source $C_i$
\beq \label{vielbeins}
e^a_\mu=\left(
\begin{tabular}{ cc}
 $ -v^j e^{a}_j$, &
  $e^a_i+C_i v^j e^a_j$
\end{tabular}
 \right),
\qquad
e^{a\mu}=\left(
\begin{tabular}{ c}
 $ C_j e^{aj}$ \\
  $e^{ai}$
\end{tabular}
 \right).
\eeq
One can easily check that $e^a_\mu=g^{\text{nc}}_{\mu\nu}e^{a\nu}$ and $e^{a\mu}=g_{\text{nc}}^{\mu\nu}e^{a}_{\nu}$. In addition, $g^{\text{nc}}_{\mu\nu}=e^a_\mu e^a_\nu$ and $g_{\text{nc}}^{\mu\nu}=e^{a\mu} e^{a\nu}$. This form is also consistent with the orthogonality requirements $e^a_\mu V^\mu=0$, $n_\mu e^{a\mu}=0$. One can thus interpret $V^\mu$ and $n_\mu$ as vielbein vector and one-form with $a=0$.

Given Eq. \eqref{omega}, we find
\begin{widetext}
\beq
\begin{split}
\omega_\mu&=\frac 1 2 \epsilon^{ab} e^{a\nu}\big[\partial_\mu e^b_\nu-\underbrace{V^\lambda \partial_\mu n_\nu e^b_\lambda}_{=0}-\frac{1}{2}g_{\text{nc}}^{\lambda \rho} \left( \partial_\mu g^{\text{nc}}_{\nu\rho}+\partial_\nu g^{\text{nc}}_{\mu \rho}-\partial_\rho g^{\text{nc}}_{\mu\nu}\right) e^b_\lambda \big] \\
&=\frac 1 2 \epsilon^{ab} e^{a\nu}\Big[\partial_\mu e^b_\nu-\frac{1}{2} \left( \partial_\mu g^{\text{nc}}_{\nu\rho}+\partial_\nu g^{\text{nc}}_{\mu \rho}-\partial_\rho g^{\text{nc}}_{\mu\nu}\right) e^{b\rho} \Big] \\
&=\frac 1 2 \epsilon^{ab} e^{a \nu}\partial_\mu e^b_\nu
-\frac{1}{2} \epsilon^{ab} e^{a\nu} e^{b\rho} \partial_\nu g^{\text{nc}}_{\mu \rho}.
\end{split}
\eeq
Written in components
\beq
\begin{split}
\omega_t=\frac{1}{2}\Big(&\epsilon^{ab}e^{aj} \partial_t e^{b}_{j}+\varepsilon^{ij} \Big[\partial_i v_j + \partial_t(C_i v_j) 
\Big] \Big), \\
\omega_i=\frac{1}{2}\Big(&\epsilon^{ab}e^{aj} \partial_i e^{b}_{j}-\varepsilon^{jk}\left[\partial_j g_{ik}+v_j\partial_i C_k+\partial_j(v_k C_i)+v^2 C_i \partial_j C_k \right] - \\
& \varepsilon^{jk}C_j \left[\partial_t g_{ik}+\partial_t(v_k C_i)+v^2 C_i \partial_t C_k \right] \Big),
\end{split}
\eeq
where the constraint \eqref{const} was used.
In the following we will need only the terms that are linear in $C_i$, hence it is sufficient to write
\beq \label{dC}
\begin{split}
\omega_t=&\frac{1}{2}\Big(\epsilon^{ab}e^{aj} \partial_t e^{b}_{j}+\varepsilon^{ij} \Big[\partial_i v_j +\partial_t(C_i  v_j) \Big] \Big), \\
\omega_i=&\frac{1}{2}\Big(\epsilon^{ab}e^{aj} \partial_i e^{b}_{j}-\varepsilon^{jk}\left[\partial_j g_{ik}+v_j\partial_i C_k+\partial_j(v_k C_i) +C_j \partial_t g_{ik} \right] \Big)+\mathcal{O}(C^2).
\end{split}
\eeq
\end{widetext}
The expression \eqref{spincon} is recovered when $v_i$ is equal to the superfluid velocity. This can be seen as a gauge fixing of the Newton-Cartan geometry.
\subsection{Covariant description of superfluids}

We are now in position to write the action of the nonrelativistic superfluid in covariant form in the Newton-Cartan formalism. For the conventional superfluid we find
\beq \label{ncaction}
S =  \int\! dt d\mathbf{x}\,\sqrt \gamma \left[\rho V^\mu (\d_\mu\theta - \tilde A_\mu)
      -\epsilon(\rho) \right],
\eeq
where we introduced the superfluid density $\rho$ that transforms as a scalar, i.e.,  $\delta \rho=-\xi^\kappa \partial_\kappa \rho$. In addition, we defined $\gamma_{\mu\nu}\equiv g_{\mu\nu}^{\text{nc}}+n_\mu n_\nu$ \cite{Jensen2014} with the determinant $\gamma=e^{-2\Phi}g$. Eq. \eqref{ncaction} is a generalization of our construction in \cite{Hoyos2013} to the case with nonvanishing $\Phi$ and $C_i$. For completeness, in Appendix \ref{AppA} we rewrite the theory solely in terms of the Goldstone boson field $\theta$.

It is straightforward to generalize this construction to the case of the chiral superfluid which in Newton-Cartan formalism is described by the action
\beq \label{chiralnc}
S=\int\! dt d\mathbf{x}\,\sqrt \gamma \left[\rho V^\mu (\d_\mu\theta - \tilde A_\mu-s \omega_\mu)
      -\epsilon(\rho) \right].
\eeq
In this formulation the current is convective
\beq
J^\mu\equiv-\frac{1}{\sqrt \gamma }\frac{\delta S}{\delta A_\mu}=\rho V^\mu
\eeq
which implies that in Newton-Cartan geometry the scalar $\rho$ can be constructed covariantly as
\beq
\rho=n_\mu J^\mu.
\eeq

In the special case $\Phi=C_i=0$, it is easy to solve the equations of motion for $\rho$ and $v^i$ with the result
\beq \label{eomrv1}
\begin{split}
\mathscr{D}_t \theta &=- v^i \mathscr{D}_i \theta-\frac{1}{2} g_{ij}v^i v^j
      +\epsilon'(\rho) \\
v^i&=-g^{ij}\mathscr{D}_j\theta+\frac{s}{2} \varepsilon^{ij}\partial_j \ln \rho,
\end{split}
\eeq
where $\mathscr{D}_i\theta=\partial_i \theta- A_i - s \omega_i$. Note that in Newton-Cartan formalism the superfluid velocity $v^i$ is given by (minus) the covariant derivative of the Goldstone field plus an additional term that is proportional and perpendicular to the gradient of the superfluid density.\footnote{Since the superfluid velocity has no unique definition in the microscopic theory, its redefinition is allowed and is known as the frame transformation in the theory of hydrodynamics \cite{Kovtun:2012rj}.} This term is responsible for the edge part of the current that appears in the ground state in the presence of inhomogeneties and gives rise to the angular momentum \eqref{angmom}. Also due to this extra term, in the present formulation one finds  $T^{ij}=T^{ij}_{\text{\text{ideal}}}+T^{ij}_{\text{Hall}}$.

Finally, we will generalize the dual description of a superfluid presented in Sec. \ref{dualsec} to the covariant form in Newton-Cartan spacetime.
First, from Eqs. \eqref{ncaction} and \eqref{chiralnc} we notice that for $\Phi\ne 0$ the conservation equation of particle number is given by
\beq
\partial_\mu \left(\sqrt{\gamma} J^\mu \right)=0,
\eeq
which is identically satisfied by
\beq
J^\mu=\varepsilon_\text{nc}^{\mu\nu\rho}\partial_\nu a_\rho,
\eeq
where we introduced $\varepsilon_\text{nc}^{\mu\nu\rho} \equiv \sqrt{\gamma}^{-1}\epsilon^{\mu\nu\rho}$. In Newton-Cartan spacetime $\varepsilon_\text{nc}^{\mu\nu\rho}$ transforms as a tensor.\footnote{This can be demonstrated by using the identity $\varepsilon_\text{nc}^{\mu\nu\rho} \partial_\kappa \xi^\kappa=\varepsilon_\text{nc}^{\mu\nu\kappa}\partial_\kappa \xi^\rho+\epsilon_\text{nc}^{\mu\kappa\rho}\partial_\kappa \xi^\nu+ \varepsilon_\text{nc}^{\kappa \nu \rho} \partial_\kappa \xi^{\mu}$.}
Given this tensor and the current $J^{\mu}$ that transforms as a vector
\beq
\delta J^{\mu}=-\xi^\kappa \partial_\kappa J^{\mu}+J^\kappa \partial_k \xi^\mu,
\eeq
the gauge potential must transform simply as a one-form
\beq
\delta a_{\mu}=-\xi^\kappa \partial_\kappa a_{\mu}-a_\kappa \partial_\mu \xi^\kappa.
\eeq

The dual theory of the conventional superfluid in Newton-Cartan spacetime is given by the action
\beq
S=\int dt d\mathbf{x} \sqrt{\gamma} \mathcal{L}_{sf}
\eeq
with the Lagrangian
\beq
\mathcal{L}_{sf}=-\epsilon(\rho)- \varepsilon_\text{nc}^{\mu\nu\rho}\tilde A_{\mu}\partial_{\nu}a_{\rho},
\eeq
where $\rho=n_\mu J^\mu=\varepsilon_\text{nc}^{\mu\nu\rho}n_\mu \partial_\nu a_\rho$ or using the parametrization \eqref{nV} one finds $\rho=b+\varepsilon^{ij} C_i e_j$. This Lagrangian follows directly from Eq. \eqref{ncaction}.
Using the equation of motion
\beq
v^i=-\frac{\varepsilon^{ij}e_j}{\rho}
\eeq
we can  eliminate the velocity field $v^i$ and rewrite the Lagrangian as
\beq
\mathcal{L}_{sf}=e^{2\Phi} \frac{g^{ij}e_i e_j}{2\rho}-\epsilon(\rho)- e^\Phi \varepsilon^{\mu\nu\rho} A_{\mu}\partial_{\nu}a_{\rho},
\eeq
which is the generalization of Eq. \eqref{sfL} to the case with nonvanishing $\Phi$ and $C_i$.

The covariant form of the dual theory of a chiral superfluid in Newton-Cartan geometry is given by
\beq
\begin{split}
\mathcal{L}_{ch}&=\mathcal{L}_{sf}+\mathcal{L}_{WZ} \\
&=-\epsilon(\rho)- \varepsilon_\text{nc}^{\mu\nu\rho}\tilde A_{\mu}\partial_{\nu}a_{\rho}-\varepsilon_\text{nc}^{\mu\nu\rho}\omega_{\mu}\partial_{\nu}a_{\rho}.
\end{split}
\eeq

\section{Energy current} \label{Encur}
Provided the background sources are \emph{static}, the system has time translation symmetry. By Noether theorem this leads to the conservation of the energy current $J^\mu_\epsilon$. The Newton-Cartan formalism developed above is a convenient framework for the calculation of $J^\mu_\epsilon$.  The current is defined by
\beq
\delta S=\int dt d \mathbf{x} \sqrt{\gamma} J^\mu_\epsilon \partial_\mu \xi^t,
\eeq
which follows from the invariance of the effective action under (global) time translations.
Employing now Eq. \eqref{new-gci} we find
\beq \label{encurrent}
\begin{split}
J^t_\epsilon&=\frac{1}{\sqrt{g}e^{-\Phi}}\left(\frac{\delta S}{\delta \Phi}-\frac{\delta S}{\delta A_0}A_0+\frac{\delta S}{\delta C_i}C_i \right)\to \frac{1}{\sqrt{g}}\frac{\delta S}{\delta \Phi}, \\
J^i_\epsilon&=\frac{1}{\sqrt{g}e^{-\Phi}}\left(\frac{\delta S}{\delta C_i}-\frac{\delta S}{\delta A_i}A_0 \right)\to \frac{1}{\sqrt{g}}\frac{\delta S}{\delta C_i},
\end{split}
\eeq
where the most right expressions are valid provided $A_0=\Phi=C_i=0$. This explains why $\Phi$ and $C_i$ serve as external sources for the energy density and current, respectively.

By applying now the prescription \eqref{encurrent} to the action \eqref{ncaction} we first calculate the energy current of the conventional superfluid. For $A_0=\Phi=C_i=0$ one finds
\beq \label{encurrid}
\begin{split}
J^t_{\epsilon, \text{\text{ideal}}}&=\frac{1}{\sqrt{g}}\frac{\delta S}{\delta \Phi}=\frac 12 \rho g_{kl} v^k v^l +\epsilon(\rho), \\
J^i_{\epsilon, \text{\text{ideal}}}&=\frac{1}{\sqrt{g}}\frac{\delta S}{\delta C_i}=\rho \mathscr{D}_t\theta v^i \\
&=\left(P+\epsilon(\rho)+\frac 12 \rho g_{kl} v^k v^l \right) v^i,
\end{split}
\eeq
which is the well-known result for an ideal fluid. In the second equation we used the equations of motion \eqref{eomrv1} with $s=0$ and the relation $P+\epsilon(\rho)=\rho \epsilon'(\rho)$.

Now we are ready to calculate how the energy current \eqref{encurrid} is modified in the chiral superfluid. For simplicity we will only consider the background $A_0=\Phi=C_i=0$. Since
\beq
S_{WZ}=\int dt d \mathbf{x} \sqrt{\gamma} \mathscr{L}_{WZ}
\eeq
does not depend on $\Phi$ , we find
\beq
J^t_\epsilon=\frac{1}{\sqrt{g}}\frac{\delta S}{\delta \Phi}=\frac 12 \rho g_{kl} v^k v^l +\epsilon(\rho).
\eeq
Hence there is no correction to the energy density compared to the conventional superfluid. On the other hand, the modification of the spatial energy current is nontrivial. Indeed, for the chiral superfluid we find
\beq \label{I}
\delta S=\int\! dt d\mathbf{x}\,\sqrt g \rho \left[\delta C_l v^l \mathscr{D}_t \theta-s\delta \omega_t-s v^l \delta \omega_l \right]
\eeq
with
\beq \label{II}
\begin{split}
\delta \omega_t=&\frac{1}{2}\varepsilon^{ij}\partial_t(\delta C_i  v_j), \\
\delta \omega_i=&-\frac{1}{2}\varepsilon^{jk}\left[v_j\partial_i \delta C_k+\partial_j(v_k \delta C_i) +\delta C_j \partial_t g_{ik} \right],
\end{split}
\eeq
where Eq. \eqref{dC} was applied. In Appendix \ref{AppE} the resulting energy current is found to be given by
\beq \label{encurrentp}
\begin{split}
J^l_\epsilon=&\frac{1}{\sqrt{g}}\frac{\delta S}{\delta C_i} \\ =&J^l_{\epsilon, \text{\text{ideal}}}+
\frac{s}{2}\Big[\frac{1}{\sqrt{g}}\partial_t (\sqrt{g} \rho) +\nabla_i(\rho v^i)\Big]\varepsilon^{lj}v_j+T_{\text{Hall}}^{lm} v_m.
\end{split}
\eeq
The second term vanishes if the equations of motion are used. The last term is the correction due to the modification of the stress tensor. A similar correction also arises in dissipative Navier-Stokes hydrodynamics \cite{LL6} and parity-violating hydrodynamics of normal fluids \cite{Kaminski2014}.

\section{Conclusion}

In this paper we have constructed the leading order terms in the effective action of conventional and chiral two-dimensional fermionic superfluids using the dual gauge field formulation.
A similar low-energy description for superconductors was developed in \cite{Hansson2004,Hansson2012,Hansson2013}, with the important difference that for the superfluid the action is not purely topological due to the presence of a gapless Goldstone mode.

Compared to other works, we impose nonrelativistic diffeomorphism invariance \cite{Son:2005rv} that puts stringent constraints on the form of the effective action. It also allows us to consider superfluids living on curved manifolds. We use the Newton-Cartan formalism  \cite{Son2013,Christensen:2013lma,*Christensen:2013rfa,Geracie2014,Gromov2014,Bradlyn2014, Banerjee:2014nja,*Banerjee:2014pya,Brauner2014a} to present a covariant formulation of the superfluid with sources that couple to all conserved currents. We have also shown that the parity-breaking relativistic superfluid of \cite{GRS} reduces to the chiral superfluid in the non-relativistic limit. In particular the coupling of the dual gauge field to the Euler topological current reduces to the famous Wen-Zee term.

Even though the chiral superfluid studied here has Galilean invariance, the Newton-Cartan formalism can be applied to more general cases without Galilean or Lorentzian invariance \cite{Christensen:2013lma,*Christensen:2013rfa,Gromov2014,Bradlyn2014, Brauner2014a}. In particular, it would be useful to construct a covariant effective action of chiral superconductors, i.e, charged superfluids coupled to a dynamical electromagnetic field.


\section*{Acknowledgments:}
We would like to acknowledge discussions with Dam Thanh Son which inspired this work.
We are also thankful to Alexander Abanov, Tom\'a\v{s} Brauner, Siavash Golkar, Andrey Gromov, Thors Hans Hansson, Hitoshi Murayama, Haruki Watanabe  for useful discussions.  This work
was supported by  US DOE Grant
No.\ DE-FG02-97ER41014  and by the I-CORE program of Planning and
Budgeting Committee and the Israeli Science Foundation (grant number 1937/12).
\appendix
\section{Goldstone boson formulation} \label{AppA}
In this Appendix we demonstrate that the theory defined by the Lagrangian \eqref{HydroLag} is equivalent to the leading-order general coordinate invariant theory of nonrelativistic s-wave superfluid found previously in \cite{Greiter:1989qb,Son:2005rv}. Up to a surface term the Lagrangian \eqref{HydroLag} is
\beq \label{A1}
\mathcal{L}_{sf}=\rho \mathscr{D}_t\theta  +\rho v^i \mathscr{D}_i\theta
      + \frac 12 \rho g_{ij} v^i v^j - \epsilon(\rho),
\eeq
where we introduce the covariant derivative $\mathscr{D}_\mu \theta=\partial_\mu \theta-A_\mu$.

The Euler-Lagrange equation $\delta S/ \delta v^i=0$ gives us
\beq \label{A2}
v_i=-\mathscr{D}_i \theta
\eeq
and thus $\theta$ can be identified as the Goldstone field of the spontaneously broken $U(1)_N$ particle number symmetry. Now we can substitute Eq. \eqref{A2} into Eq. \eqref{A1} and find
\beq
\mathcal{L}_{sf}=\rho\underbrace{\left( \mathscr{D}_t \theta -\frac{g^{ij}}{2} \mathscr{D}_i \theta\mathscr{D}_j \theta \right)}_{X}-\epsilon(\rho),
\eeq
where we introduced the combination $X$ which is general coordinate invariant and reduces to the chemical potential in the ground state. Due to the equation of motion $\delta S/ \delta \rho=0$, $\rho$ and $X$ are the Legendre-dual variables and we finally arrive at \cite{Greiter:1989qb,Son:2005rv}
\beq \label{PX}
\mathscr{L}_{sf}=P(X),
\eeq
where $P=\rho d\epsilon/d\rho-\epsilon$ is the thermodynamic pressure as the function of the generalized chemical potential $X$.

The same derivation can be repeated in the presence of the sources $\Phi$ and $C_i$ (see Sec. \ref{ncG}). Starting from Eq. \eqref{ncaction} we once again obtain Eq. \eqref{PX} with
\beq \label{Xs}
X=D_t \theta -\frac{g^{ij}}{2} D_i \theta D_j \theta,
\eeq
where the modified covariant derivatives are
\beq
\begin{split}
D_t \theta&\equiv e^\Phi \mathscr{D}_t \theta, \\
D_i \theta&\equiv  \mathscr{D}_i \theta +C_i \mathscr{D}_t \theta.
\end{split}
\eeq
In Newton-Cartan spacetime the general coordinate invariant \eqref{Xs} can be conveniently written as
\beq
\begin{split}
X&=V^\mu \tilde {\mathscr{D}}_\mu \theta-\frac{1}{2}e^{a \mu} \tilde {\mathscr{D}}_\mu \theta e^{a \nu} \tilde {\mathscr{D}}_\nu \theta \\
&=V^\mu \tilde {\mathscr{D}}_\mu \theta-\frac{1}{2}g_{\text{nc}}^{\mu\nu} \tilde {\mathscr{D}}_\mu \theta \tilde {\mathscr{D}}_\nu \theta,
\end{split}
\eeq
where $ \tilde {\mathscr{D}}_\mu \theta \equiv \mathscr{D}_\mu \theta-\tilde A_\mu$ with $\tilde A_\mu$ defined by Eqs. \eqref{Atilde}. This form was also found in \cite{Brauner2014a}.

\section{Linearized hydrodynamics in dual language} \label{AppB}
Consider small phonon fluctuations around the homogeneous superfluid ground state in flat space with $A_\mu=0$.
In the dual picture the vacuum of this theory has $b_{\text{GS}}=\rho_{\text{GS}}$ and $\mathbf{e}_{\text{GS}}=0$, which follows from Eq. \eqref{rhov}.
By expanding Eq. \eqref{sfL} to the quadratic order in fluctuations $\delta b= b-b_{\text{GS}}$ and $\mathbf{e}$ we find
\beq
\begin{split}
\mathcal{L}_{sf}&=\frac{1}{2 \rho_{\text{GS}}}\mathbf{e}^2-\epsilon_{\text{GS}}-\epsilon'_{\text{GS}} \delta b-\frac{1}{2}\epsilon''_{\text{GS}} \delta b^2 \\
&\to \frac{1}{2\rho_{\text{GS}}}\mathbf{e}^2- \frac{\epsilon''_{\text{GS}}}{2} \delta b^2,
\end{split}
\eeq
where in the second line we dropped the constant and linear terms. The linearized approximation of Eq. \eqref{sfL} thus gives rise to the linear relativistic electrodynamics, where the effective speed of light is fixed by the speed of sound $c_s=\sqrt{dP/d\rho}|_{\rho=\rho_{\text{GS}}}=\sqrt{\epsilon''_{\text{GS}} \rho_{\text{GS}}}$.
\section{From vielbeins to metric} \label{AppC}
Here we express the gravitomagnetic field $B_\omega$ and gravitoelectric field $E_{\omega i}$ solely in terms of the spatial metric $g_{ij}$. First, using Eq. \eqref{spincon} and the orthonormality of the vielbein, it is straightforward to show that
\beq
B_\omega=\varepsilon^{ij}\partial_i \omega_j=R/2,
\eeq
where the Ricci scalar $R=g^{ij}R_{ij}$ with
\beq
\begin{split}
 R_{ij} &= \partial_k \Gamma^k_{ij}
  -\partial_i \Gamma^k_{j k}
  +\Gamma^k_{ij}\Gamma^l_{k l}
  - \Gamma^k_{il}\Gamma^l_{jk}, \\
  \Gamma^i_{jk} &= \frac12 g^{il}(\partial_j g_{lk}
  + \partial_k g_{lj} - \partial_l g_{jk}).
\end{split}
\eeq
On the other hand, using Eq. \eqref{spincon} the gravitoelectric field can be written as
\beq
E_{\omega i}=\frac{1}{2}\left[\underbrace{\epsilon^{ab}(\partial_t e^{aj}\partial_i e^b_j-\partial_i e^{aj}\partial_t e^b_j)}_{M_i}-\partial_t(\varepsilon^{jk}\partial_j g_{ik})-\partial_i B \right].
\eeq
It is convenient to express $M_i$ as
\beq
M_i= \epsilon^{ab}\Big\{\partial_t e^{aj}[\nabla_i e^b_j+\Gamma^k_{ij}e^b_k]-[\nabla_i e^{aj}-\Gamma^j_{ik} e^{ak}]\partial_t e^b_j\Big\}.
\eeq
Using $\nabla_i e^b_j=-\omega_i \epsilon^{bc}e^c_j$, it is now easy to show that the terms with covariant derivatives cancel and we end up with
\beq
M_i=\Gamma^k_{ij}\partial_t (\varepsilon^{jl} g_{lk}).
\eeq
The electric field thus reads
\beq
E_{\omega i}=\frac{1}{2}\left[\Gamma^k_{ij}\partial_t (\varepsilon^{jl} g_{kl})-\partial_t(\varepsilon^{jl}\partial_j g_{il})-\partial_i B \right],
\eeq
which using
\beq
\varepsilon^{jl}\nabla_j g_{il}=\varepsilon^{jl}\partial_j g_{il}-\varepsilon^{jl}\Gamma^k_{ji}g_{kl}=0
\eeq
can be simplified to
\beq
E_{\omega i}=\frac{1}{2}\left[-\partial_t (\Gamma^k_{ij})\varepsilon^{jl} g_{kl}-\partial_i B \right].
\eeq
\section{Nonrelativistic limit of Euler current} \label{AppD}
Here we provide some details on the calculation of the nonrelativistic limit of the Euler current $\mR{J}^\mu$.

First we note that since
\begin{equation}
\nabla_i \mR{u}_j \sim \frac{1}{c},\ \ \nabla_0 \mR{u}_i\sim \nabla_i \mR{u}_0\sim \frac{1}{c^2}, \ \ \nabla_0 \mR{u}_0 \sim \frac{1}{c^3},
\end{equation}
we can neglect in the Euler current \eqref{Euler} the terms depending on the velocities. In addition, since
\begin{equation}
\varepsilon^{\alpha\beta\gamma}\mR{u}_\alpha R_{\nu\lambda\beta\gamma}\simeq -\varepsilon^{ij} R_{\nu\lambda ij}-2\varepsilon^{ij}\frac{v_i}{c}R_{\nu\lambda 0j}
\end{equation}
the second term is $O(1/c^2)$ compared with the first one and can thus be neglected. Hence we find
\begin{equation}
8\pi \mR{J}^\mu\simeq \frac{1}{2}\varepsilon^{\mu\nu\lambda}\varepsilon^{ij}R_{\nu\lambda ij}.
\end{equation}
As a result, the time component of the Euler current equals to the spatial scalar curvature
\begin{equation}
8\pi\mR{J}^0= \frac{1}{2}\varepsilon^{kl}\varepsilon^{ij}R_{kl ij}=\frac{1}{2} (g^{ki}g^{jl}-g^{kj} g^{il})R_{kl ij}=R=2 B_\omega.
\end{equation}
The spatial part $\mR{J}^k$ can be expressed using the mixed components of the Ricci tensor
\begin{equation}
8\pi \mR{J}^k\simeq-\varepsilon^{kl}\varepsilon^{ij}R_{0l ij}=-(g^{ki}g^{jl}-g^{kj} g^{il})R_{0l ij}.
\end{equation}
We can now use that in three dimensions
\begin{equation}\label{riemann}
R_{\mu\nu\rho\lambda}=2(g_{\mu[\rho} R_{\lambda]\nu}-g_{\nu[\rho} R_{\lambda]\mu})-R g_{\mu[\rho}g_{\lambda]\nu}
\end{equation}
and up to the relativistic corrections
\beq
R_{ij}\simeq \frac{1}{2}R g_{ij},
\eeq
which leads to
\begin{equation}
R_{0l ij} \simeq g_{lj} R_{0i}-g_{li}R_{0j}.
\end{equation}
This implies
\begin{equation} \label{Ei}
8\pi \mR{J}^k\simeq -2 g^{ik}R_{0i}.
\end{equation}
The Ricci tensor is given by
\begin{equation} \label{Ricci}
\begin{split}
R_{0i}&\simeq \partial_k \Gamma^k_{i0}-\partial_i \Gamma^k_{0k}+\Gamma^k_{kl}\Gamma^l_{0i}-\Gamma^l_{ik}\Gamma^k_{0l} \\
& =\nabla_k \Gamma^k_{0i}-\partial_i \Gamma^k_{0k},
\end{split}
\end{equation}
where the covariant derivative is defined with respect to the spatial metric only.
Using the metric \eqref{metric} one finds in the nonrelativistic limit \cite{Son:2005rv}
\begin{equation} \label{Gamma}
\Gamma^i_{0j}\simeq \frac{1}{2c}\left(g^{ik}\dot{g}_{kj}+g^{ik}{\cal F}_{kj} \right).
\end{equation}
With the help of the last equation, the Ricci tensor \eqref{Ricci}  can alternatively be written as
\begin{equation}
\begin{split}
R_{0i}&\simeq\frac{1}{2c}\nabla_k\left(g^{kl}\dot{g}_{li}+\varepsilon^k_{\ i} {\cal B} \right)-\frac{1}{2c}\partial_i\left( g^{kl}\dot{g}_{kl}\right) \\
&=\frac{1}{2c}\nabla_k\left(g^{kl}\dot{g}_{li}+\varepsilon^k_{\ i} {\cal B} \right)-\frac{1}{2c}\nabla_i\left( g^{kl}\dot{g}_{kl}\right)\\
&=\frac{1}{2c}\left(g^{kl}\nabla_k \dot{g}_{li}+\varepsilon^k_{\ i} \partial_k{\cal B} \right)-\frac{1}{2c}g^{kl}\nabla_i\dot{g}_{kl}.
\end{split}
\end{equation}
Putting this into \eqref{Ei} and using
\begin{equation}
\nabla_i \dot{g}_{jk}=\dot\Gamma^l_{ij}g_{lk} +\dot\Gamma^l_{ik}g_{jl}
\end{equation}
we find
\begin{equation}
\begin{split}
8\pi \mR{J}^k&\simeq-\frac 1 c \big[\varepsilon^{nk}\partial_n {\cal B}+g^{jk}g^{nl}\left(\nabla_n \dot{g}_{lj}-\nabla_j\dot{g}_{nl}\right) \big]  \\
&=-\frac 1 c \big[\varepsilon^{nk}\partial_n {\cal B}+g^{jk}g^{nl}\left(\dot\Gamma^s_{nl}g_{sj} -\dot\Gamma^s_{jl}g_{ns}\right) \big]\\
&=-\frac 1 c \big[\varepsilon^{nk}\partial_n {\cal B}+(g^{nl}g^{kj}-g^{kl}g^{nj})g_{sj}\dot\Gamma^s_{nl}\big] \\
&=-\frac 1 c \varepsilon^{nk} \big[\partial_n {\cal B}+\varepsilon^{lj}g_{sj}\dot\Gamma^s_{nl}\big] \\
&=-\frac{2}{c}\varepsilon^{kn}[E_{\omega\,n}+\partial_n \mR{B}/2],
\end{split}
\end{equation}
where $\mR{B}=\varepsilon^{ij}\partial_i \mR{A}_j$.
\begin{widetext}
\section{Energy current calculation} \label{AppE}
In this Appendix we present how to compute the energy current $J^l_\epsilon$.
First, using Eqs. \eqref{I}, \eqref{II} we find
\beq
J^l_\epsilon=\frac{1}{\sqrt{g}}\frac{\delta S}{\delta C_l}=\rho \mathscr{D}_t \theta v^l+\frac{s}{2}\Big(\partial_t \rho \varepsilon^{lj}v_j+\varepsilon^{lj}\partial_i(\rho v^i v_j)-\varepsilon^{jk}\partial_j(\rho v^l)v_k+\varepsilon^{lk}\rho v^i \partial_t g_{ik} \Big).
\eeq
Given the equation of motion \eqref{eomrv1}, we can eliminate $\mathscr{D}_t \theta$ with the result
\beq \label{jint}
\begin{split}
J^l_\epsilon&=J^l_{\epsilon, \text{ideal}}-\frac s 2 v^l v_i  \varepsilon^{ij} \partial_j \rho+\frac{s}{2}\Big(\partial_t \rho \varepsilon^{lj}v_j+\varepsilon^{lj}\partial_i(\rho v^i v_j)-\varepsilon^{jk}\partial_j(\rho v^l)v_k+\varepsilon^{lk}\rho v^i \partial_t g_{ik} \Big) \\
=&J^l_{\epsilon, \text{ideal}}+\frac{s}{2}\Big(\partial_t \rho \varepsilon^{lj}v_j+\varepsilon^{lj}\partial_i(\rho v^i v_j)-\varepsilon^{jk} \rho\partial_j v^l v_k+\varepsilon^{lk}\rho v^i \partial_t g_{ik} \Big) \\
=&J^l_{\epsilon, \text{ideal}}+\frac{s}{2}\Big([\partial_t \rho +\partial_i(\rho v^i)]\varepsilon^{lj}v_j+\varepsilon^{lj}\rho v^i\partial_i v_j-\varepsilon^{jk} \rho\partial_j v^l v_k+\varepsilon^{lk}\rho v^i \partial_t g_{ik} \Big) \\
=&J^l_{\epsilon, \text{ideal}}+\frac{s}{2}\Big[\frac{1}{\sqrt{g}}\partial_t (\sqrt{g} \rho) +\nabla_i(\rho v^i)\Big]\varepsilon^{lj}v_j+ \Delta J^l_\epsilon
\end{split}
\eeq
with
\beq \label{DeltaJ}
 \Delta J^l_\epsilon=\frac {s \rho} {2} \Big(\varepsilon^{lj} v^i\partial_i v_j-\varepsilon^{jk}g^{ln} \partial_jv_n v_k-\varepsilon^{jk} \partial_j g^{ln} v_n v_k-\frac{1}{2} v^m g^{ij}\partial_m g_{ij}\varepsilon^{lk}v_k+\varepsilon^{lk} v^i \partial_t g_{ik}-\frac{1}{2} g^{ij}\partial_t g_{ij}\varepsilon^{lk}v_k \Big).
\eeq
One can check that $\Delta J^l_\epsilon$ agrees with the covariant expression
\beq \label{encurrentcov}
\Delta J^l_\epsilon
=\frac s 4 \rho (\varepsilon^{lr} g^{ms}+\varepsilon^{mr} g^{ls})(\nabla_r v_s+\nabla_s v_r+\partial_t g _{rs}) v_m
=T_{\text{Hall}}^{lm} v_m. \\
\eeq
This can be achieved either by a direct comparison\footnote{Since every two-dimensional Riemannian manifold is conformally flat, it is especially convenient for the purpose of the comparison to use the coordinates, where $g_{ij}=\sqrt{g}\delta_{ij}$. } or by the following calculation: First, pick from the bracket of Eq. \eqref{DeltaJ} only the terms depending on derivatives of the velocity
\begin{equation} \label{res1}
\begin{split}
&\varepsilon^{lj} v^i\partial_i v_j-\varepsilon^{jk}g^{ln} \partial_jv_n v_k=\left[\varepsilon^{lj} g^{ni}+\varepsilon^{ni}g^{lj} \right]\partial_i v_j v_n=
\frac{1}{2}\left[\varepsilon^{lj} g^{ni}+\varepsilon^{ni}g^{lj} \right]\sigma_{ij}v_n\\
&=\frac{1}{2}\left[\varepsilon^{lj} g^{ni}+\varepsilon^{nj}g^{li} \right]\sigma_{ij}v_n=\frac{1}{2}\left[\varepsilon^{lj} g^{ni}+\varepsilon^{nj}g^{li}\right]\left(\left[\nabla_i v_j+\nabla_j v_i-\delta_{ij}\nabla_k v^k +\partial_t g_{ij}\right]v_n+2\Gamma^k_{ij}v_k v_n- \partial_t g_{ij}v_n\right)\\
&=\frac{2}{s\rho}T^{ln}_{\text{Hall}}v_n+\frac{1}{2}\left[\varepsilon^{lj} g^{ni}+\varepsilon^{nj}g^{li} \right]\left(2\Gamma^k_{ij}v_k v_n- \partial_t g_{ij}v_n\right)
\end{split}
\end{equation}
where we took advantage of the decomposition
\begin{equation}
\partial_i v_j =\frac{1}{2}\sigma_{ij}+\frac{1}{2}\omega\varepsilon_{ij}+\frac{1}{2}\theta \delta_{ij},
\end{equation}
where
\begin{equation}
\sigma_{ij}= \partial_i v_j+\partial_j v_i-\delta_{ij}\partial_k v^k, \ \ \omega=\varepsilon^{ij}\partial_i v_j, \ \ \theta=\partial_k v^k.
\end{equation}
Now substitute Eq. \eqref{res1} into Eq. \eqref{DeltaJ} and consider the difference between Eqs. \eqref{DeltaJ} and \eqref{encurrentcov}. If we collect the terms depending on the time derivatives we find
\begin{equation}
\begin{split}
&-\frac{1}{2}\left[\varepsilon^{lj} g^{ni}+\varepsilon^{ni}g^{lj} \right]\partial_t g_{ij}v_n+\varepsilon^{lk} v^i \partial_t g_{ik}-\frac{1}{2} g^{ij}\partial_t g_{ij}\varepsilon^{lk}v_k=-\frac{1}{2}\left[\varepsilon^{lj} g^{ni}+\varepsilon^{ni}g^{lj}  -2\varepsilon^{lj} g^{in}+g^{ij}\varepsilon^{ln}\right]\partial_t g_{ij} v_n\\
&=-\frac{1}{2}\left[-\varepsilon^{lj} g^{ni}+\varepsilon^{ni}g^{lj} +g^{ij}\varepsilon^{ln}\right]\partial_t g_{ij} v_n=-\frac{1}{2}\left[-\varepsilon^{li} g^{nj}+\varepsilon^{ni}g^{lj} +g^{ij}\varepsilon^{ln}\right]\partial_t g_{ij} v_n\\
&=-\frac{1}{2}\left[\varepsilon^{nl}g^{ij} +g^{ij}\varepsilon^{ln}\right]\partial_t g_{ij} v_n=0.
\end{split}
\end{equation}
Finally, the terms depending on spatial derivatives of the metric vanish as well
\begin{equation}
\begin{split}
&v_n v_k\left[\left(\varepsilon^{lj}g^{ni}+\varepsilon^{nj}g^{li}\right)\Gamma^k_{ij} +\varepsilon^{jk}\left(g^{in}\Gamma^l_{ij}+\Gamma^n_{ij}g^{il}\right)-\Gamma^j_{ij}\varepsilon^{lk} g^{in} \right]\\
&= v_n v_k \left[g^{ni}\left(\varepsilon^{lj}\Gamma^k_{ij}-\varepsilon^{kj}\Gamma^l_{ij}\right)+g^{li}\left(\varepsilon^{nj}\Gamma^k_{ij}+ \varepsilon^{jn}\Gamma^k_{ij}\right) -\Gamma^j_{ij}\varepsilon^{lk} g^{in} \right]\\
&=v_n v_k \left[g^{ni}\varepsilon^{lk}\Gamma^j_{ji} -\Gamma^j_{ij}\varepsilon^{lk} g^{in} \right]=0.
\end{split}
\end{equation}
This proves that Eqs. \eqref{DeltaJ} and \eqref{encurrentcov} are equivalent.

\end{widetext}
\bibliography{ref}

\begin{thebibliography}{53}%
\makeatletter
\providecommand \@ifxundefined [1]{%
 \@ifx{#1\undefined}
}%
\providecommand \@ifnum [1]{%
 \ifnum #1\expandafter \@firstoftwo
 \else \expandafter \@secondoftwo
 \fi
}%
\providecommand \@ifx [1]{%
 \ifx #1\expandafter \@firstoftwo
 \else \expandafter \@secondoftwo
 \fi
}%
\providecommand \natexlab [1]{#1}%
\providecommand \enquote  [1]{``#1''}%
\providecommand \bibnamefont  [1]{#1}%
\providecommand \bibfnamefont [1]{#1}%
\providecommand \citenamefont [1]{#1}%
\providecommand \href@noop [0]{\@secondoftwo}%
\providecommand \href [0]{\begingroup \@sanitize@url \@href}%
\providecommand \@href[1]{\@@startlink{#1}\@@href}%
\providecommand \@@href[1]{\endgroup#1\@@endlink}%
\providecommand \@sanitize@url [0]{\catcode `\\12\catcode `\$12\catcode
  `\&12\catcode `\#12\catcode `\^12\catcode `\_12\catcode `\%12\relax}%
\providecommand \@@startlink[1]{}%
\providecommand \@@endlink[0]{}%
\providecommand \url  [0]{\begingroup\@sanitize@url \@url }%
\providecommand \@url [1]{\endgroup\@href {#1}{\urlprefix }}%
\providecommand \urlprefix  [0]{URL }%
\providecommand \Eprint [0]{\href }%
\providecommand \doibase [0]{http://dx.doi.org/}%
\providecommand \selectlanguage [0]{\@gobble}%
\providecommand \bibinfo  [0]{\@secondoftwo}%
\providecommand \bibfield  [0]{\@secondoftwo}%
\providecommand \translation [1]{[#1]}%
\providecommand \BibitemOpen [0]{}%
\providecommand \bibitemStop [0]{}%
\providecommand \bibitemNoStop [0]{.\EOS\space}%
\providecommand \EOS [0]{\spacefactor3000\relax}%
\providecommand \BibitemShut  [1]{\csname bibitem#1\endcsname}%
\let\auto@bib@innerbib\@empty
\bibitem [{\citenamefont {Volovik}(1992)}]{volovik1992exotic}%
  \BibitemOpen
  \bibfield  {author} {\bibinfo {author} {\bibfnamefont {G.~E.}\ \bibnamefont
  {Volovik}},\ }\href@noop {} {\emph {\bibinfo {title} {Exotic properties of
  superfluid 3He}}},\ Vol.~\bibinfo {volume} {1}\ (\bibinfo  {publisher} {World
  Scientific},\ \bibinfo {year} {1992})\BibitemShut {NoStop}%
\bibitem [{\citenamefont {Vollhardt}\ and\ \citenamefont
  {Wolfle}(1990)}]{vollhardt}%
  \BibitemOpen
  \bibfield  {author} {\bibinfo {author} {\bibfnamefont {D.}~\bibnamefont
  {Vollhardt}}\ and\ \bibinfo {author} {\bibfnamefont {P.}~\bibnamefont
  {Wolfle}},\ }\href@noop {} {\emph {\bibinfo {title} {Superfluid phases of
  helium 3}}}\ (\bibinfo  {publisher} {Taylor and Francis Ltd},\ \bibinfo
  {year} {1990})\BibitemShut {NoStop}%
\bibitem [{\citenamefont {Volovik}(2009)}]{volovikbook}%
  \BibitemOpen
  \bibfield  {author} {\bibinfo {author} {\bibfnamefont {G.~E.}\ \bibnamefont
  {Volovik}},\ }\href@noop {} {\emph {\bibinfo {title} {The universe in a
  helium droplet}}},\ Vol.\ \bibinfo {volume} {117}\ (\bibinfo  {publisher}
  {Oxford University Press New York},\ \bibinfo {year} {2009})\BibitemShut
  {NoStop}%
\bibitem [{\citenamefont {Kitaev}(2003)}]{kitaev2003fault}%
  \BibitemOpen
  \bibfield  {author} {\bibinfo {author} {\bibfnamefont {A.~Y.}\ \bibnamefont
  {Kitaev}},\ }\href@noop {} {\bibfield  {journal} {\bibinfo  {journal} {Ann.
  Phys.}\ }\textbf {\bibinfo {volume} {303}},\ \bibinfo {pages} {2} (\bibinfo
  {year} {2003})}\BibitemShut {NoStop}%
\bibitem [{\citenamefont {Nayak}\ \emph {et~al.}(2008)\citenamefont {Nayak},
  \citenamefont {Simon}, \citenamefont {Stern}, \citenamefont {Freedman},\ and\
  \citenamefont {Sarma}}]{nayak2008non}%
  \BibitemOpen
  \bibfield  {author} {\bibinfo {author} {\bibfnamefont {C.}~\bibnamefont
  {Nayak}}, \bibinfo {author} {\bibfnamefont {S.~H.}\ \bibnamefont {Simon}},
  \bibinfo {author} {\bibfnamefont {A.}~\bibnamefont {Stern}}, \bibinfo
  {author} {\bibfnamefont {M.}~\bibnamefont {Freedman}}, \ and\ \bibinfo
  {author} {\bibfnamefont {S.~D.}\ \bibnamefont {Sarma}},\ }\href@noop {}
  {\bibfield  {journal} {\bibinfo  {journal} {Rev. Mod. Phys.}\ }\textbf
  {\bibinfo {volume} {80}},\ \bibinfo {pages} {1083} (\bibinfo {year}
  {2008})}\BibitemShut {NoStop}%
\bibitem [{\citenamefont {G\"unter}\ \emph {et~al.}(2005)\citenamefont
  {G\"unter}, \citenamefont {St\"oferle}, \citenamefont {Moritz}, \citenamefont
  {K\"ohl},\ and\ \citenamefont {Esslinger}}]{Gunter}%
  \BibitemOpen
  \bibfield  {author} {\bibinfo {author} {\bibfnamefont {K.}~\bibnamefont
  {G\"unter}}, \bibinfo {author} {\bibfnamefont {T.}~\bibnamefont
  {St\"oferle}}, \bibinfo {author} {\bibfnamefont {H.}~\bibnamefont {Moritz}},
  \bibinfo {author} {\bibfnamefont {M.}~\bibnamefont {K\"ohl}}, \ and\ \bibinfo
  {author} {\bibfnamefont {T.}~\bibnamefont {Esslinger}},\ }\href {\doibase
  10.1103/PhysRevLett.95.230401} {\bibfield  {journal} {\bibinfo  {journal}
  {Phys. Rev. Lett.}\ }\textbf {\bibinfo {volume} {95}},\ \bibinfo {pages}
  {230401} (\bibinfo {year} {2005})}\BibitemShut {NoStop}%
\bibitem [{\citenamefont {Read}\ and\ \citenamefont
  {Green}(2000)}]{PhysRevB.61.10267}%
  \BibitemOpen
  \bibfield  {author} {\bibinfo {author} {\bibfnamefont {N.}~\bibnamefont
  {Read}}\ and\ \bibinfo {author} {\bibfnamefont {D.}~\bibnamefont {Green}},\
  }\href {\doibase 10.1103/PhysRevB.61.10267} {\bibfield  {journal} {\bibinfo
  {journal} {Phys. Rev. B}\ }\textbf {\bibinfo {volume} {61}},\ \bibinfo
  {pages} {10267} (\bibinfo {year} {2000})}\BibitemShut {NoStop}%
\bibitem [{\citenamefont {{Fisher}}(1999)}]{Fisher}%
  \BibitemOpen
  \bibfield  {author} {\bibinfo {author} {\bibfnamefont {M.~P.~A.}\
  \bibnamefont {{Fisher}}},\ }in\ \href@noop {} {\emph {\bibinfo {booktitle}
  {Topological Aspects of Low Dimensional Systems}}},\ \bibinfo {editor}
  {edited by\ \bibinfo {editor} {\bibfnamefont {A.}~\bibnamefont {{Comtet}}},
  \bibinfo {editor} {\bibfnamefont {T.}~\bibnamefont {{Jolicoeur}}}, \bibinfo
  {editor} {\bibfnamefont {S.}~\bibnamefont {{Ouvry}}}, \ and\ \bibinfo
  {editor} {\bibfnamefont {F.}~\bibnamefont {{David}}}}\ (\bibinfo {year}
  {1999})\ p.\ \bibinfo {pages} {575},\ \Eprint
  {http://arxiv.org/abs/cond-mat/9806164} {cond-mat/9806164} \BibitemShut
  {NoStop}%
\bibitem [{\citenamefont {Son}\ and\ \citenamefont
  {Wingate}(2006)}]{Son:2005rv}%
  \BibitemOpen
  \bibfield  {author} {\bibinfo {author} {\bibfnamefont {D.}~\bibnamefont
  {Son}}\ and\ \bibinfo {author} {\bibfnamefont {M.}~\bibnamefont {Wingate}},\
  }\href {\doibase 10.1016/j.aop.2005.11.001} {\bibfield  {journal} {\bibinfo
  {journal} {Ann. Phys.}\ }\textbf {\bibinfo {volume} {321}},\ \bibinfo {pages}
  {197} (\bibinfo {year} {2006})}\BibitemShut {NoStop}%
\bibitem [{\citenamefont {Son}(2013)}]{Son2013}%
  \BibitemOpen
  \bibfield  {author} {\bibinfo {author} {\bibfnamefont {D.~T.}\ \bibnamefont
  {Son}},\ }\href@noop {} {\  (\bibinfo {year} {2013})},\ \Eprint
  {http://arxiv.org/abs/1306.0638} {arXiv:1306.0638 [cond-mat.mes-hall]}
  \BibitemShut {NoStop}%
\bibitem [{\citenamefont {Christensen}\ \emph
  {et~al.}(2014{\natexlab{a}})\citenamefont {Christensen}, \citenamefont
  {Hartong}, \citenamefont {Obers},\ and\ \citenamefont
  {Rollier}}]{Christensen:2013lma}%
  \BibitemOpen
  \bibfield  {author} {\bibinfo {author} {\bibfnamefont {M.~H.}\ \bibnamefont
  {Christensen}}, \bibinfo {author} {\bibfnamefont {J.}~\bibnamefont
  {Hartong}}, \bibinfo {author} {\bibfnamefont {N.~A.}\ \bibnamefont {Obers}},
  \ and\ \bibinfo {author} {\bibfnamefont {B.}~\bibnamefont {Rollier}},\ }\href
  {\doibase 10.1103/PhysRevD.89.061901} {\bibfield  {journal} {\bibinfo
  {journal} {Phys.Rev.}\ }\textbf {\bibinfo {volume} {D89}},\ \bibinfo {pages}
  {061901} (\bibinfo {year} {2014}{\natexlab{a}})},\ \Eprint
  {http://arxiv.org/abs/1311.4794} {arXiv:1311.4794 [hep-th]} \BibitemShut
  {NoStop}%
\bibitem [{\citenamefont {Christensen}\ \emph
  {et~al.}(2014{\natexlab{b}})\citenamefont {Christensen}, \citenamefont
  {Hartong}, \citenamefont {Obers},\ and\ \citenamefont
  {Rollier}}]{Christensen:2013rfa}%
  \BibitemOpen
  \bibfield  {author} {\bibinfo {author} {\bibfnamefont {M.~H.}\ \bibnamefont
  {Christensen}}, \bibinfo {author} {\bibfnamefont {J.}~\bibnamefont
  {Hartong}}, \bibinfo {author} {\bibfnamefont {N.~A.}\ \bibnamefont {Obers}},
  \ and\ \bibinfo {author} {\bibfnamefont {B.}~\bibnamefont {Rollier}},\ }\href
  {\doibase 10.1007/JHEP01(2014)057} {\bibfield  {journal} {\bibinfo  {journal}
  {JHEP}\ }\textbf {\bibinfo {volume} {1401}},\ \bibinfo {pages} {057}
  (\bibinfo {year} {2014}{\natexlab{b}})},\ \Eprint
  {http://arxiv.org/abs/1311.6471} {arXiv:1311.6471 [hep-th]} \BibitemShut
  {NoStop}%
\bibitem [{\citenamefont {Geracie}\ \emph {et~al.}(2014)\citenamefont
  {Geracie}, \citenamefont {Son}, \citenamefont {Wu},\ and\ \citenamefont
  {Wu}}]{Geracie2014}%
  \BibitemOpen
  \bibfield  {author} {\bibinfo {author} {\bibfnamefont {M.}~\bibnamefont
  {Geracie}}, \bibinfo {author} {\bibfnamefont {D.~T.}\ \bibnamefont {Son}},
  \bibinfo {author} {\bibfnamefont {C.}~\bibnamefont {Wu}}, \ and\ \bibinfo
  {author} {\bibfnamefont {S.-F.}\ \bibnamefont {Wu}},\ }\href@noop {} {\
  (\bibinfo {year} {2014})},\ \Eprint {http://arxiv.org/abs/1407.1252}
  {arXiv:1407.1252 [cond-mat.mes-hall]} \BibitemShut {NoStop}%
\bibitem [{\citenamefont {{Gromov}}\ and\ \citenamefont
  {{Abanov}}(2015)}]{Gromov2014}%
  \BibitemOpen
  \bibfield  {author} {\bibinfo {author} {\bibfnamefont {A.}~\bibnamefont
  {{Gromov}}}\ and\ \bibinfo {author} {\bibfnamefont {A.~G.}\ \bibnamefont
  {{Abanov}}},\ }\href@noop {} {\bibfield  {journal} {\bibinfo  {journal}
  {Phys. Rev. Lett.}\ }\textbf {\bibinfo {volume} {114}},\ \bibinfo {pages}
  {016802} (\bibinfo {year} {2015})},\ \Eprint {http://arxiv.org/abs/1407.2908}
  {arXiv:1407.2908 [cond-mat.str-el]} \BibitemShut {NoStop}%
\bibitem [{\citenamefont {{Bradlyn}}\ and\ \citenamefont
  {{Read}}(2014)}]{Bradlyn2014}%
  \BibitemOpen
  \bibfield  {author} {\bibinfo {author} {\bibfnamefont {B.}~\bibnamefont
  {{Bradlyn}}}\ and\ \bibinfo {author} {\bibfnamefont {N.}~\bibnamefont
  {{Read}}},\ }\href@noop {} {\  (\bibinfo {year} {2014})},\ \Eprint
  {http://arxiv.org/abs/1407.2911} {arXiv:1407.2911 [cond-mat.mes-hall]}
  \BibitemShut {NoStop}%
\bibitem [{\citenamefont {Banerjee}\ \emph {et~al.}(2015)\citenamefont
  {Banerjee}, \citenamefont {Mitra},\ and\ \citenamefont
  {Mukherjee}}]{Banerjee:2014nja}%
  \BibitemOpen
  \bibfield  {author} {\bibinfo {author} {\bibfnamefont {R.}~\bibnamefont
  {Banerjee}}, \bibinfo {author} {\bibfnamefont {A.}~\bibnamefont {Mitra}}, \
  and\ \bibinfo {author} {\bibfnamefont {P.}~\bibnamefont {Mukherjee}},\
  }\href@noop {} {\bibfield  {journal} {\bibinfo  {journal}
  {Class.Quant.Grav.}\ }\textbf {\bibinfo {volume} {32}},\ \bibinfo {pages}
  {045010} (\bibinfo {year} {2015})},\ \Eprint {http://arxiv.org/abs/1407.3617}
  {arXiv:1407.3617 [hep-th]} \BibitemShut {NoStop}%
\bibitem [{\citenamefont {Banerjee}\ \emph {et~al.}(2014)\citenamefont
  {Banerjee}, \citenamefont {Mitra},\ and\ \citenamefont
  {Mukherjee}}]{Banerjee:2014pya}%
  \BibitemOpen
  \bibfield  {author} {\bibinfo {author} {\bibfnamefont {R.}~\bibnamefont
  {Banerjee}}, \bibinfo {author} {\bibfnamefont {A.}~\bibnamefont {Mitra}}, \
  and\ \bibinfo {author} {\bibfnamefont {P.}~\bibnamefont {Mukherjee}},\
  }\href@noop {} {\bibfield  {journal} {\bibinfo  {journal} {Phys. Lett. B}\
  }\textbf {\bibinfo {volume} {737}},\ \bibinfo {pages} {369} (\bibinfo {year}
  {2014})},\ \Eprint {http://arxiv.org/abs/1404.4491} {arXiv:1404.4491 [gr-qc]}
  \BibitemShut {NoStop}%
\bibitem [{\citenamefont {Brauner}\ \emph {et~al.}(2014)\citenamefont
  {Brauner}, \citenamefont {Endlich}, \citenamefont {Monin},\ and\
  \citenamefont {Penco}}]{Brauner2014a}%
  \BibitemOpen
  \bibfield  {author} {\bibinfo {author} {\bibfnamefont {T.}~\bibnamefont
  {Brauner}}, \bibinfo {author} {\bibfnamefont {S.}~\bibnamefont {Endlich}},
  \bibinfo {author} {\bibfnamefont {A.}~\bibnamefont {Monin}}, \ and\ \bibinfo
  {author} {\bibfnamefont {R.}~\bibnamefont {Penco}},\ }\href@noop {}
  {\bibfield  {journal} {\bibinfo  {journal} {Phys. Rev. D}\ }\textbf {\bibinfo
  {volume} {90}},\ \bibinfo {pages} {105016} (\bibinfo {year} {2014})},\
  \Eprint {http://arxiv.org/abs/1407.7730} {arXiv:1407.7730 [hep-th]}
  \BibitemShut {NoStop}%
\bibitem [{\citenamefont {Stone}\ and\ \citenamefont {Roy}(2004)}]{Stone2004}%
  \BibitemOpen
  \bibfield  {author} {\bibinfo {author} {\bibfnamefont {M.}~\bibnamefont
  {Stone}}\ and\ \bibinfo {author} {\bibfnamefont {R.}~\bibnamefont {Roy}},\
  }\href {\doibase 10.1103/PhysRevB.69.184511} {\bibfield  {journal} {\bibinfo
  {journal} {Phys. Rev. B}\ }\textbf {\bibinfo {volume} {69}},\ \bibinfo
  {pages} {184511} (\bibinfo {year} {2004})}\BibitemShut {NoStop}%
\bibitem [{\citenamefont {Hoyos}\ \emph {et~al.}(2014)\citenamefont {Hoyos},
  \citenamefont {Moroz},\ and\ \citenamefont {Son}}]{Hoyos2013}%
  \BibitemOpen
  \bibfield  {author} {\bibinfo {author} {\bibfnamefont {C.}~\bibnamefont
  {Hoyos}}, \bibinfo {author} {\bibfnamefont {S.}~\bibnamefont {Moroz}}, \ and\
  \bibinfo {author} {\bibfnamefont {D.~T.}\ \bibnamefont {Son}},\ }\href
  {\doibase 10.1103/PhysRevB.89.174507} {\bibfield  {journal} {\bibinfo
  {journal} {Phys. Rev. B}\ }\textbf {\bibinfo {volume} {89}},\ \bibinfo
  {pages} {174507} (\bibinfo {year} {2014})},\ \Eprint
  {http://arxiv.org/abs/1305.3925} {arXiv:1305.3925 [cond-mat.quant-gas]}
  \BibitemShut {NoStop}%
\bibitem [{\citenamefont {Zakharov}\ and\ \citenamefont
  {Kuznetsov}(1997)}]{Zakharov1997}%
  \BibitemOpen
  \bibfield  {author} {\bibinfo {author} {\bibfnamefont {V.~E.}\ \bibnamefont
  {Zakharov}}\ and\ \bibinfo {author} {\bibfnamefont {E.~A.}\ \bibnamefont
  {Kuznetsov}},\ }\href@noop {} {\bibfield  {journal} {\bibinfo  {journal}
  {Phys. Usp.}\ }\textbf {\bibinfo {volume} {40}},\ \bibinfo {pages} {1087}
  (\bibinfo {year} {1997})}\BibitemShut {NoStop}%
\bibitem [{\citenamefont {{Schakel}}(1998)}]{1998cond.mat..5152S}%
  \BibitemOpen
  \bibfield  {author} {\bibinfo {author} {\bibfnamefont {A.~M.~J.}\
  \bibnamefont {{Schakel}}},\ }\href@noop {} {\  (\bibinfo {year} {1998})},\
  \Eprint {http://arxiv.org/abs/cond-mat/9805152} {cond-mat/9805152}
  \BibitemShut {NoStop}%
\bibitem [{\citenamefont {Son}(2007)}]{Son:2007}%
  \BibitemOpen
  \bibfield  {author} {\bibinfo {author} {\bibfnamefont {D.~T.}\ \bibnamefont
  {Son}},\ }\href {\doibase 10.1103/PhysRevLett.98.020604} {\bibfield
  {journal} {\bibinfo  {journal} {Phys. Rev. Lett.}\ }\textbf {\bibinfo
  {volume} {98}},\ \bibinfo {pages} {020604} (\bibinfo {year}
  {2007})}\BibitemShut {NoStop}%
\bibitem [{\citenamefont {Deser}\ \emph {et~al.}(1982)\citenamefont {Deser},
  \citenamefont {Jackiw},\ and\ \citenamefont {Templeton}}]{Deser1982}%
  \BibitemOpen
  \bibfield  {author} {\bibinfo {author} {\bibfnamefont {S.}~\bibnamefont
  {Deser}}, \bibinfo {author} {\bibfnamefont {R.}~\bibnamefont {Jackiw}}, \
  and\ \bibinfo {author} {\bibfnamefont {S.}~\bibnamefont {Templeton}},\
  }\href@noop {} {\bibfield  {journal} {\bibinfo  {journal} {Ann. Phys.}\
  }\textbf {\bibinfo {volume} {140}},\ \bibinfo {pages} {372} (\bibinfo {year}
  {1982})}\BibitemShut {NoStop}%
\bibitem [{\citenamefont {Brauner}\ and\ \citenamefont
  {Watanabe}(2014)}]{Brauner2014}%
  \BibitemOpen
  \bibfield  {author} {\bibinfo {author} {\bibfnamefont {T.}~\bibnamefont
  {Brauner}}\ and\ \bibinfo {author} {\bibfnamefont {H.}~\bibnamefont
  {Watanabe}},\ }\href {\doibase 10.1103/PhysRevD.89.085004} {\bibfield
  {journal} {\bibinfo  {journal} {Phys. Rev. D}\ }\textbf {\bibinfo {volume}
  {89}},\ \bibinfo {pages} {085004} (\bibinfo {year} {2014})}\BibitemShut
  {NoStop}%
\bibitem [{\citenamefont {Kovner}\ \emph {et~al.}(1991)\citenamefont {Kovner},
  \citenamefont {Rosenstein},\ and\ \citenamefont {Eliezer}}]{Kovner1991}%
  \BibitemOpen
  \bibfield  {author} {\bibinfo {author} {\bibfnamefont {A.}~\bibnamefont
  {Kovner}}, \bibinfo {author} {\bibfnamefont {B.}~\bibnamefont {Rosenstein}},
  \ and\ \bibinfo {author} {\bibfnamefont {D.}~\bibnamefont {Eliezer}},\
  }\href@noop {} {\bibfield  {journal} {\bibinfo  {journal} {Nucl. Phys. B}\
  }\textbf {\bibinfo {volume} {350}},\ \bibinfo {pages} {325} (\bibinfo {year}
  {1991})}\BibitemShut {NoStop}%
\bibitem [{\citenamefont {Janiszewski}\ and\ \citenamefont
  {Karch}(2013)}]{Janiszewski:2012nb}%
  \BibitemOpen
  \bibfield  {author} {\bibinfo {author} {\bibfnamefont {S.}~\bibnamefont
  {Janiszewski}}\ and\ \bibinfo {author} {\bibfnamefont {A.}~\bibnamefont
  {Karch}},\ }\href {\doibase 10.1007/JHEP02(2013)123} {\bibfield  {journal}
  {\bibinfo  {journal} {JHEP}\ }\textbf {\bibinfo {volume} {1302}},\ \bibinfo
  {pages} {123} (\bibinfo {year} {2013})},\ \Eprint
  {http://arxiv.org/abs/1211.0005} {arXiv:1211.0005 [hep-th]} \BibitemShut
  {NoStop}%
\bibitem [{\citenamefont {Wen}\ and\ \citenamefont {Zee}(1992)}]{Wen1992}%
  \BibitemOpen
  \bibfield  {author} {\bibinfo {author} {\bibfnamefont {X.}~\bibnamefont
  {Wen}}\ and\ \bibinfo {author} {\bibfnamefont {A.}~\bibnamefont {Zee}},\
  }\href@noop {} {\bibfield  {journal} {\bibinfo  {journal} {Phys. Rev. Lett.}\
  }\textbf {\bibinfo {volume} {69}},\ \bibinfo {pages} {953} (\bibinfo {year}
  {1992})}\BibitemShut {NoStop}%
\bibitem [{\citenamefont {Greiter}\ \emph {et~al.}(1989)\citenamefont
  {Greiter}, \citenamefont {Wilczek},\ and\ \citenamefont
  {Witten}}]{Greiter:1989qb}%
  \BibitemOpen
  \bibfield  {author} {\bibinfo {author} {\bibfnamefont {M.}~\bibnamefont
  {Greiter}}, \bibinfo {author} {\bibfnamefont {F.}~\bibnamefont {Wilczek}}, \
  and\ \bibinfo {author} {\bibfnamefont {E.}~\bibnamefont {Witten}},\ }\href
  {\doibase 10.1142/S0217984989001400} {\bibfield  {journal} {\bibinfo
  {journal} {Mod. Phys. Lett. B}\ }\textbf {\bibinfo {volume} {3}},\ \bibinfo
  {pages} {903} (\bibinfo {year} {1989})}\BibitemShut {NoStop}%
\bibitem [{\citenamefont {{Tada}}\ \emph {et~al.}(2014)\citenamefont {{Tada}},
  \citenamefont {{Nie}},\ and\ \citenamefont {{Oshikawa}}}]{Tada2014}%
  \BibitemOpen
  \bibfield  {author} {\bibinfo {author} {\bibfnamefont {Y.}~\bibnamefont
  {{Tada}}}, \bibinfo {author} {\bibfnamefont {W.}~\bibnamefont {{Nie}}}, \
  and\ \bibinfo {author} {\bibfnamefont {M.}~\bibnamefont {{Oshikawa}}},\
  }\href@noop {} {\bibfield  {journal} {\bibinfo  {journal} {ArXiv e-prints}\ }
  (\bibinfo {year} {2014})},\ \Eprint {http://arxiv.org/abs/1409.7459}
  {arXiv:1409.7459 [cond-mat.supr-con]} \BibitemShut {NoStop}%
\bibitem [{\citenamefont {{Volovik}}(2014)}]{Volovik2014}%
  \BibitemOpen
  \bibfield  {author} {\bibinfo {author} {\bibfnamefont {G.~E.}\ \bibnamefont
  {{Volovik}}},\ }\href@noop {} {\bibfield  {journal} {\bibinfo  {journal}
  {JETP Lett.}\ }\textbf {\bibinfo {volume} {100}},\ \bibinfo {pages} {843}
  (\bibinfo {year} {2014})},\ \Eprint {http://arxiv.org/abs/1409.8638}
  {arXiv:1409.8638 [cond-mat.str-el]} \BibitemShut {NoStop}%
\bibitem [{\citenamefont {{Huang}}\ \emph {et~al.}(2014)\citenamefont
  {{Huang}}, \citenamefont {{Taylor}},\ and\ \citenamefont
  {{Kallin}}}]{Huang2014}%
  \BibitemOpen
  \bibfield  {author} {\bibinfo {author} {\bibfnamefont {W.}~\bibnamefont
  {{Huang}}}, \bibinfo {author} {\bibfnamefont {E.}~\bibnamefont {{Taylor}}}, \
  and\ \bibinfo {author} {\bibfnamefont {C.}~\bibnamefont {{Kallin}}},\
  }\href@noop {} {\bibfield  {journal} {\bibinfo  {journal} {Phys. Rev. B}\
  }\textbf {\bibinfo {volume} {90}},\ \bibinfo {pages} {224519} (\bibinfo
  {year} {2014})},\ \Eprint {http://arxiv.org/abs/1410.0377} {arXiv:1410.0377
  [cond-mat.supr-con]} \BibitemShut {NoStop}%
\bibitem [{\citenamefont {Hoyos}\ and\ \citenamefont
  {Son}(2012)}]{Hoyos:2011ez}%
  \BibitemOpen
  \bibfield  {author} {\bibinfo {author} {\bibfnamefont {C.}~\bibnamefont
  {Hoyos}}\ and\ \bibinfo {author} {\bibfnamefont {D.~T.}\ \bibnamefont
  {Son}},\ }\href@noop {} {\bibfield  {journal} {\bibinfo  {journal} {Phys.
  Rev. Lett.}\ }\textbf {\bibinfo {volume} {108}},\ \bibinfo {pages} {066805}
  (\bibinfo {year} {2012})},\ \Eprint {http://arxiv.org/abs/1109.2651}
  {arXiv:1109.2651 [cond-mat.mes-hall]} \BibitemShut {NoStop}%
\bibitem [{\citenamefont {{Gromov}}\ and\ \citenamefont
  {{Abanov}}(2014)}]{2014arXiv1403.5809G}%
  \BibitemOpen
  \bibfield  {author} {\bibinfo {author} {\bibfnamefont {A.}~\bibnamefont
  {{Gromov}}}\ and\ \bibinfo {author} {\bibfnamefont {A.~G.}\ \bibnamefont
  {{Abanov}}},\ }\href@noop {} {\bibfield  {journal} {\bibinfo  {journal}
  {Phys. Rev. Lett.}\ }\textbf {\bibinfo {volume} {113}},\ \bibinfo {pages}
  {266802} (\bibinfo {year} {2014})},\ \Eprint {http://arxiv.org/abs/1403.5809}
  {arXiv:1403.5809 [cond-mat.str-el]} \BibitemShut {NoStop}%
\bibitem [{\citenamefont {Mermin}\ and\ \citenamefont
  {Muzikar}(1980)}]{Mermin1980}%
  \BibitemOpen
  \bibfield  {author} {\bibinfo {author} {\bibfnamefont {N.~D.}\ \bibnamefont
  {Mermin}}\ and\ \bibinfo {author} {\bibfnamefont {P.}~\bibnamefont
  {Muzikar}},\ }\href {\doibase 10.1103/PhysRevB.21.980} {\bibfield  {journal}
  {\bibinfo  {journal} {Phys. Rev. B}\ }\textbf {\bibinfo {volume} {21}},\
  \bibinfo {pages} {980} (\bibinfo {year} {1980})}\BibitemShut {NoStop}%
\bibitem [{\citenamefont {Avron}\ \emph {et~al.}(1995)\citenamefont {Avron},
  \citenamefont {Seiler},\ and\ \citenamefont {Zograf}}]{Avron:1995fg}%
  \BibitemOpen
  \bibfield  {author} {\bibinfo {author} {\bibfnamefont {J.}~\bibnamefont
  {Avron}}, \bibinfo {author} {\bibfnamefont {R.}~\bibnamefont {Seiler}}, \
  and\ \bibinfo {author} {\bibfnamefont {P.}~\bibnamefont {Zograf}},\ }\href
  {\doibase 10.1103/PhysRevLett.75.697} {\bibfield  {journal} {\bibinfo
  {journal} {Phys.Rev.Lett.}\ }\textbf {\bibinfo {volume} {75}},\ \bibinfo
  {pages} {697} (\bibinfo {year} {1995})}\BibitemShut {NoStop}%
\bibitem [{\citenamefont {{Avron}}(1997)}]{1997physics..12050A}%
  \BibitemOpen
  \bibfield  {author} {\bibinfo {author} {\bibfnamefont {J.~E.}\ \bibnamefont
  {{Avron}}},\ }\href@noop {} {\  (\bibinfo {year} {1997})},\ \Eprint
  {http://arxiv.org/abs/arXiv:physics/9712050} {arXiv:physics/9712050}
  \BibitemShut {NoStop}%
\bibitem [{\citenamefont {Hoyos}(2014)}]{Hoyos2014}%
  \BibitemOpen
  \bibfield  {author} {\bibinfo {author} {\bibfnamefont {C.}~\bibnamefont
  {Hoyos}},\ }\href {\doibase 10.1142/S0217979214300072} {\bibfield  {journal}
  {\bibinfo  {journal} {Int. J. Mod. Phys.}\ }\textbf {\bibinfo {volume}
  {B28}},\ \bibinfo {pages} {1430007} (\bibinfo {year} {2014})},\ \Eprint
  {http://arxiv.org/abs/1403.4739} {arXiv:1403.4739 [cond-mat.mes-hall]}
  \BibitemShut {NoStop}%
\bibitem [{\citenamefont {Wilczek}\ and\ \citenamefont
  {Zee}(1983)}]{Wilczek1983}%
  \BibitemOpen
  \bibfield  {author} {\bibinfo {author} {\bibfnamefont {F.}~\bibnamefont
  {Wilczek}}\ and\ \bibinfo {author} {\bibfnamefont {A.}~\bibnamefont {Zee}},\
  }\href@noop {} {\bibfield  {journal} {\bibinfo  {journal} {Phys. Rev. Lett.}\
  }\textbf {\bibinfo {volume} {51}},\ \bibinfo {pages} {2250} (\bibinfo {year}
  {1983})}\BibitemShut {NoStop}%
\bibitem [{\citenamefont {Haldane}\ and\ \citenamefont
  {Wu}(1985)}]{Haldane1985}%
  \BibitemOpen
  \bibfield  {author} {\bibinfo {author} {\bibfnamefont {F.~D.~M.}\
  \bibnamefont {Haldane}}\ and\ \bibinfo {author} {\bibfnamefont {Y.-S.}\
  \bibnamefont {Wu}},\ }\href {\doibase 10.1103/PhysRevLett.55.2887} {\bibfield
   {journal} {\bibinfo  {journal} {Phys. Rev. Lett.}\ }\textbf {\bibinfo
  {volume} {55}},\ \bibinfo {pages} {2887} (\bibinfo {year}
  {1985})}\BibitemShut {NoStop}%
\bibitem [{\citenamefont {{Lee}}\ and\ \citenamefont
  {{Fisher}}(1991)}]{Lee1991}%
  \BibitemOpen
  \bibfield  {author} {\bibinfo {author} {\bibfnamefont {D.-H.}\ \bibnamefont
  {{Lee}}}\ and\ \bibinfo {author} {\bibfnamefont {M.~P.~A.}\ \bibnamefont
  {{Fisher}}},\ }\href {\doibase 10.1142/S0217979291001061} {\bibfield
  {journal} {\bibinfo  {journal} {Int. J. Mod. Phys. B}\ }\textbf {\bibinfo
  {volume} {5}},\ \bibinfo {pages} {2675} (\bibinfo {year} {1991})}\BibitemShut
  {NoStop}%
\bibitem [{\citenamefont {{Ariad}}\ \emph {et~al.}(2014)\citenamefont
  {{Ariad}}, \citenamefont {{Seradjeh}},\ and\ \citenamefont
  {{Grosfeld}}}]{Ariad2014}%
  \BibitemOpen
  \bibfield  {author} {\bibinfo {author} {\bibfnamefont {D.}~\bibnamefont
  {{Ariad}}}, \bibinfo {author} {\bibfnamefont {B.}~\bibnamefont {{Seradjeh}}},
  \ and\ \bibinfo {author} {\bibfnamefont {E.}~\bibnamefont {{Grosfeld}}},\
  }\href@noop {} {\  (\bibinfo {year} {2014})},\ \Eprint
  {http://arxiv.org/abs/1407.2553} {arXiv:1407.2553 [cond-mat.supr-con]}
  \BibitemShut {NoStop}%
\bibitem [{\citenamefont {{Hansson}}\ \emph {et~al.}(2012)\citenamefont
  {{Hansson}}, \citenamefont {{Karlhede}},\ and\ \citenamefont
  {{Sato}}}]{Hansson2012}%
  \BibitemOpen
  \bibfield  {author} {\bibinfo {author} {\bibfnamefont {T.~H.}\ \bibnamefont
  {{Hansson}}}, \bibinfo {author} {\bibfnamefont {A.}~\bibnamefont
  {{Karlhede}}}, \ and\ \bibinfo {author} {\bibfnamefont {M.}~\bibnamefont
  {{Sato}}},\ }\href {\doibase 10.1088/1367-2630/14/6/063017} {\bibfield
  {journal} {\bibinfo  {journal} {New J. Phys.}\ }\textbf {\bibinfo {volume}
  {14}},\ \bibinfo {eid} {063017} (\bibinfo {year} {2012})},\ \Eprint
  {http://arxiv.org/abs/1105.5031} {arXiv:1105.5031 [cond-mat.supr-con]}
  \BibitemShut {NoStop}%
\bibitem [{\citenamefont {{Hansson}}\ \emph {et~al.}(2013)\citenamefont
  {{Hansson}}, \citenamefont {{Kvorning}},\ and\ \citenamefont {{Parameswaran
  Nair}}}]{Hansson2013}%
  \BibitemOpen
  \bibfield  {author} {\bibinfo {author} {\bibfnamefont {T.~H.}\ \bibnamefont
  {{Hansson}}}, \bibinfo {author} {\bibfnamefont {T.}~\bibnamefont
  {{Kvorning}}}, \ and\ \bibinfo {author} {\bibfnamefont {V.}~\bibnamefont
  {{Parameswaran Nair}}},\ }\href@noop {} {\  (\bibinfo {year} {2013})},\
  \Eprint {http://arxiv.org/abs/1310.8284} {arXiv:1310.8284 [cond-mat.str-el]}
  \BibitemShut {NoStop}%
\bibitem [{\citenamefont {Golkar}\ \emph {et~al.}(2014)\citenamefont {Golkar},
  \citenamefont {Roberts},\ and\ \citenamefont {Son}}]{GRS}%
  \BibitemOpen
  \bibfield  {author} {\bibinfo {author} {\bibfnamefont {S.}~\bibnamefont
  {Golkar}}, \bibinfo {author} {\bibfnamefont {M.~M.}\ \bibnamefont {Roberts}},
  \ and\ \bibinfo {author} {\bibfnamefont {D.~T.}\ \bibnamefont {Son}},\
  }\href@noop {} {\  (\bibinfo {year} {2014})},\ \Eprint
  {http://arxiv.org/abs/1407.7540} {arXiv:1407.7540 [hep-th]} \BibitemShut
  {NoStop}%
\bibitem [{\citenamefont {Andreev}\ \emph {et~al.}(2014)\citenamefont
  {Andreev}, \citenamefont {Haack},\ and\ \citenamefont
  {Hofmann}}]{Andreev:2013qsa}%
  \BibitemOpen
  \bibfield  {author} {\bibinfo {author} {\bibfnamefont {O.}~\bibnamefont
  {Andreev}}, \bibinfo {author} {\bibfnamefont {M.}~\bibnamefont {Haack}}, \
  and\ \bibinfo {author} {\bibfnamefont {S.}~\bibnamefont {Hofmann}},\ }\href
  {\doibase 10.1103/PhysRevD.89.064012} {\bibfield  {journal} {\bibinfo
  {journal} {Phys.Rev.}\ }\textbf {\bibinfo {volume} {D89}},\ \bibinfo {pages}
  {064012} (\bibinfo {year} {2014})},\ \Eprint {http://arxiv.org/abs/1309.7231}
  {arXiv:1309.7231 [hep-th]} \BibitemShut {NoStop}%
\bibitem [{\citenamefont {Andreev}(2015)}]{Andreev:2014gia}%
  \BibitemOpen
  \bibfield  {author} {\bibinfo {author} {\bibfnamefont {O.}~\bibnamefont
  {Andreev}},\ }\href@noop {} {\bibfield  {journal} {\bibinfo  {journal} {Phys.
  Rev. D}\ }\textbf {\bibinfo {volume} {91}},\ \bibinfo {pages} {024035}
  (\bibinfo {year} {2015})},\ \Eprint {http://arxiv.org/abs/1408.7031}
  {arXiv:1408.7031 [hep-th]} \BibitemShut {NoStop}%
\bibitem [{\citenamefont {Son}(2008)}]{Son2008}%
  \BibitemOpen
  \bibfield  {author} {\bibinfo {author} {\bibfnamefont {D.~T.}\ \bibnamefont
  {Son}},\ }\href@noop {} {\bibfield  {journal} {\bibinfo  {journal} {Phys.
  Rev. D}\ }\textbf {\bibinfo {volume} {78}},\ \bibinfo {pages} {046003}
  (\bibinfo {year} {2008})}\BibitemShut {NoStop}%
\bibitem [{\citenamefont {Jensen}(2014)}]{Jensen2014}%
  \BibitemOpen
  \bibfield  {author} {\bibinfo {author} {\bibfnamefont {K.}~\bibnamefont
  {Jensen}},\ }\href@noop {} {\bibfield  {journal} {\bibinfo  {journal}
  {arXiv:1408.6855 [hep-th].}\ } (\bibinfo {year} {2014})}\BibitemShut
  {NoStop}%
\bibitem [{\citenamefont {Kovtun}(2012)}]{Kovtun:2012rj}%
  \BibitemOpen
  \bibfield  {author} {\bibinfo {author} {\bibfnamefont {P.}~\bibnamefont
  {Kovtun}},\ }\href@noop {} {\  (\bibinfo {year} {2012})},\ \Eprint
  {http://arxiv.org/abs/1205.5040} {arXiv:1205.5040 [hep-th]} \BibitemShut
  {NoStop}%
\bibitem [{\citenamefont {Landau}\ and\ \citenamefont {Lifshitz}(1987)}]{LL6}%
  \BibitemOpen
  \bibfield  {author} {\bibinfo {author} {\bibfnamefont {L.~D.}\ \bibnamefont
  {Landau}}\ and\ \bibinfo {author} {\bibfnamefont {E.~M.}\ \bibnamefont
  {Lifshitz}},\ }\href@noop {} {\emph {\bibinfo {title} {Fluid Mechanics}}}\
  (\bibinfo  {publisher} {Pergamon Press},\ \bibinfo {year} {1987})\BibitemShut
  {NoStop}%
\bibitem [{\citenamefont {Kaminski}\ and\ \citenamefont
  {Moroz}(2014)}]{Kaminski2014}%
  \BibitemOpen
  \bibfield  {author} {\bibinfo {author} {\bibfnamefont {M.}~\bibnamefont
  {Kaminski}}\ and\ \bibinfo {author} {\bibfnamefont {S.}~\bibnamefont
  {Moroz}},\ }\href {\doibase 10.1103/PhysRevB.89.115418} {\bibfield  {journal}
  {\bibinfo  {journal} {Phys. Rev. B}\ }\textbf {\bibinfo {volume} {89}},\
  \bibinfo {pages} {115418} (\bibinfo {year} {2014})}\BibitemShut {NoStop}%
\bibitem [{\citenamefont {{Hansson}}\ \emph {et~al.}(2004)\citenamefont
  {{Hansson}}, \citenamefont {{Oganesyan}},\ and\ \citenamefont
  {{Sondhi}}}]{Hansson2004}%
  \BibitemOpen
  \bibfield  {author} {\bibinfo {author} {\bibfnamefont {T.~H.}\ \bibnamefont
  {{Hansson}}}, \bibinfo {author} {\bibfnamefont {V.}~\bibnamefont
  {{Oganesyan}}}, \ and\ \bibinfo {author} {\bibfnamefont {S.~L.}\ \bibnamefont
  {{Sondhi}}},\ }\href {\doibase 10.1016/j.aop.2004.05.006} {\bibfield
  {journal} {\bibinfo  {journal} {Ann. Phys.}\ }\textbf {\bibinfo {volume}
  {313}},\ \bibinfo {pages} {497} (\bibinfo {year} {2004})},\ \Eprint
  {http://arxiv.org/abs/cond-mat/0404327} {cond-mat/0404327} \BibitemShut
  {NoStop}%
\end{thebibliography}%

\end{document}